\documentclass{aa}
\usepackage{graphicx,amssymb}
\usepackage{natbib}
\usepackage[normalem]{ulem}
\bibpunct{(}{)}{;}{a}{}{,}

\DeclareRobustCommand{\ion}[2]{%
\relax\ifmmode
\ifx\testbx\f@series
{\mathbf{#1\,\mathsc{#2}}}\else
{\mathrm{#1\,\mathsc{#2}}}\fi
\else\textup{#1\,{\mdseries\textsc{#2}}}%
\fi}

\def\apjl{ApJ}%
\def\apjs{ApJS}%
\def\ao{Appl.~Opt.}%
\def\aap{A\&A}%
\newcommand{\wav}[1]{$\lambda#1\,{\rm cm}$} 
\newcommand{\Bt}{\,B_\mathrm{tot}}  
\newcommand{\Br}{\,B_{\rm reg}}  
\newcommand{\Brpa}{\,B_{{\rm reg}\parallel}}  
\newcommand{\Brpe}{\,B_{{\rm reg}\perp}}  
\newcommand{\Bur}{\,B_{r}}  
\newcommand{\But}{\,B_{\theta}}   
\newcommand{\Buz}{\,B_{z}}  
\newcommand{\bt}{\,B_{\rm tur}}  

\newcommand{\FRM}{\,{\rm rad\,m^{-2}}}

\newcommand{\kms}{\,{\rm km\,s^{-1}}}

\newcommand{\kpc}{\,{\rm kpc}}

\newcommand{\radm}{\,{\rm rad\,m^{-2}}}

\newcommand{\RM}{\,\mathrm{RM}}
\newcommand{\RMi}{\,\mathrm{RM_i}}
\newcommand{\RMfg}{\,\mathrm{RM_{fg}}}

\newcommand{\disk}{_\mathrm{d}}

\newcommand{\pb}{\,p_\mathrm{B}}
\newcommand{\pa}{\,p_\mathrm{a}}

\title{High-resolution radio continuum survey of M33\\III. Magnetic fields }

\author{ F. S. Tabatabaei\inst{1}, M. Krause\inst{1}, A. Fletcher\inst{2}, and R. Beck\inst{1} }
\institute{Max-Planck Institut f\"ur Radioastronomie, Auf dem H\"ugel 69, 53121 Bonn, Germany
\and School of Mathematics and Statistics, Newcastle University, Newcastle upon Tyne, NE1 7RU, U.K. 
}
\offprints{F. Tabatabaei\\ tabataba@mpifr-bonn.mpg.de}
\begin{document}

\titlerunning{ High-resolution radio continuum survey of M33}
\authorrunning{Tabatabaei et al.}
\abstract
{}
{We study the magnetic field structure, strength, and energy density in the Scd galaxy M33.}
{Using the linearly polarized intensity and polarization angle data at 3.6, 6.2 and 20\,cm, we determine variations of Faraday rotation and depolarization across M33. We fit a 3-D model of the regular magnetic field to the observed azimuthal distribution of polarization angles. We also analyze the spatial variation of depolarization across the galaxy.}
{Faraday rotation, measured between 3.6 and 6.2\,cm at an angular resolution of 3$\arcmin$ (0.7\,kpc), shows more variation in the south than in the north of the galaxy.  About 10\% of the nonthermal emission from M33 at 3.6\,cm is polarized. High degrees of polarization of the synchrotron emission ($> 20\%$) and strong regular magnetic fields in the sky plane ($\simeq 6.6\,\mu$G) exist in-between two northern spiral arms. We estimate the average total and regular magnetic field strengths in M33 as $\simeq$ 6.4 and 2.5\,$\mu$G, respectively. Under the assumption that the disk of M33 is flat, the regular magnetic field consists of horizontal and vertical components: however the inferred vertical field may be partly due to a galactic warp. The horizontal field is represented by an axisymmetric ($m=0$) mode from 1 to 3\,kpc radius and a superposition of axisymmetric and bisymmetric ($m=0+1$) modes from 3 to 5\,kpc radius.}
{An excess of differential Faraday rotation in the southern half together with strong Faraday dispersion in the southern spiral arms seem to be responsible for the north-south asymmetry in the observed wavelength dependent depolarization.  The presence of an axisymmetric $m=0$ mode of the regular magnetic field in each ring suggests that a galactic dynamo is operating in M33. The pitch angles of the spiral regular magnetic field are generally smaller than the pitch angles of the optical spiral arms but are twice as big as simple estimates based on the mean-field dynamo theory and M33's rotation curve. Generation of interstellar magnetic fields from turbulent gas motions in M33 is indicated by the equipartition of turbulent and magnetic energy densities.
}
\maketitle

\section{Introduction}
\label{sec:intro}

Magnetic fields in galaxies can be traced by radio polarization measurements.  The linearly polarized intensity gives information about the magnetic field strength and orientation in the plane of the sky, Faraday rotation measurements enable us to determine the magnetic field strength and direction along the line of sight and depolarizing effects can be sensitive to both components.

M33, the nearest Scd galaxy at a distance of 840\,kpc, with its large angular size \citep[high-frequency radio continuum emission was detected in an area of $35\arcmin \times 40\arcmin$, ][]{Tabatabaei_2_07} and medium inclination of 56$^{\circ}$, allows determination of the magnetic field components both parallel and perpendicular to the line of sight equally well. 
The first RM study of M33, based on polarization observations at 11.1 and 21.1\,cm \citep{Beck_79}, suggested a bisymmetric regular magnetic field structure in the disk of M33. \cite{Buczilowski_etal_91} confirmed the presence of this bisymmetric field using two further polarization maps at 6.3 and 17.4\,cm. However, these results may be affected by weak polarized intensity and the consequent high uncertainty in RM in the southern half of M33, due to the low-resolution (7.7$\arcmin$ or 1.8\,kpc) and low-sensitivity of the observations. 

Our recent observations of this galaxy provided high-sensitivity and high-resolution ($3\arcmin\simeq 0.7\kpc$) maps of total power and linearly polarized intensity at 3.6, 6.2, and 20\,cm presented by \citep{Tabatabaei_2_07}. These data are ideal to study the rotation measure (RM), magnetic fields (structure and strength), and depolarization effects in detail.

\citep{Tabatabaei_2_07} found a north-south asymmetry in the polarization distribution that is wavelength-dependent, indicating a possible north-south asymmetry in Faraday depolarization. Investigation of this possibility requires a knowledge of the distribution of RM, turbulent magnetic field and thermal electron density in the galaxy. Furthermore, depolarization is best quantified using the nonthermal degree of polarization rather than the fraction of the total radio emission that is polarized. \cite{Tabatabaei_3_07} developed a new method to separate the thermal and nonthermal components of the radio continuum emission from M33 which yielded  maps of the nonthermal synchrotron emission and the synchrotron spectral index variations across the galaxy \citep{Tabatabaei_4_07}. The nonthermal maps are used in this paper to determine the nonthermal degree of polarization at different wavelengths, and hence to study how the radio continuum emission from different parts of M33 is depolarized. 

By combining an analysis of multi-wavelength polarization angles with modeling of the wavelength dependent depolarization, \cite{Fletcher_04} and \cite{Berkhuijsen_97} derived the 3-D regular magnetic field structures in M31 and M51, respectively.
The high sensitivity of our new observations allows a similar study for M33.

We first determine the nonthermal degree of polarization using the new nonthermal maps (Sect.~\ref{sec:pol}). Then we calculate a map of the RM intrinsic to M33  with a spatial resolution of 3$\arcmin$ or 0.7\,kpc and probe its mean value in rings in the galactic plane in Sect.~\ref{sec:RM}. The regular magnetic field structure is derived in Sect.~\ref{sec:mag} and the estimated strengths are presented. We derive a map for the observed depolarization and discuss the possible  physical causes of depolarization sources in Sect.~\ref{sec:depol}. Furthermore, we discuss the resulting vertical fields and pitch angles in Sect.~6. The estimated energy density of the magnetic field is compared to the thermal and turbulent energy densities of the interstellar gas.  

\begin{table}
\begin{center}
\label{table:flux}
\caption{Integrated nonthermal flux densities and average nonthermal degree of polarization at 180$\arcsec$.}
\begin{tabular}{ l l l l } 
\hline
$\lambda$ &  S$_{\rm nth}$ &  S$_{\rm PI}$  & $\bar{{\rm P}}_{\rm nth}$ \\
(cm) &  (mJy) & (mJy) & $\%$  \\
\hline 
\hline
3.6 & 370\,$\pm$\,60    & 38\,$\pm$\,4 & 10.3\,$\pm$\,2.0  \\
6.2  & 696\,$\pm$\,110   &  79\,$\pm$\,5 & 11.3\,$\pm$\,1.9  \\
20  & 1740\,$\pm$\,65 & 115\,$\pm$\,10 &  6.6\,$\pm$\,0.6  \\
\hline
\end{tabular}
\end{center}
\end{table}

\section{Nonthermal degree of polarization}
\label{sec:pol}

The degree to which synchrotron emission is polarized reflects the degree of coherent structure in the magnetic field within one beam-area; a purely regular magnetic field will produce about 75\%\footnote{If a nonthermal spectral index of $\alpha_n=\,1$ is used.} fractional polarization of synchrotron emission. The quantity of interest is the degree of polarization of the synchrotron emission or `nonthermal degree of polarization', P$_{\rm nth} = {\rm PI}/I_{\rm nth}$, where PI is the intensity of the linearly polarized emission and $I_{\rm nth}$ is the intensity of the nonthermal emission. 

Since we observe the total intensity $I$, which consists of both nonthermal $I_{\rm nth}$ \emph{and} thermal $I_{\rm th}$ intensities at cm wavelengths, P$_{\rm nth}$ cannot be calculated straightforwardly. To date, P$_\mathrm{nth}$ has been estimated by assuming either a fixed ratio $I_\mathrm{th}/I$ or $I_\mathrm{nth}$ has been derived assuming a fixed spectral index of the synchrotron emission. Here we use a new, more robust method for determining the distribution of $I_\mathrm{nth}$ by correcting H$\alpha$ maps for dust extinction, using multi-frequency infra-red maps at the same resolution, and thus independently estimating $I_\mathrm{th}$ \citep{Tabatabaei_3_07}. Using the PI maps of \citet{Tabatabaei_2_07} and the nonthermal maps obtained by \citet{Tabatabaei_3_07}, we derived maps of P$_{\rm nth}$ at different wavelengths. 

\begin{figure}
\begin{center}
\resizebox{7cm}{!}{\includegraphics*{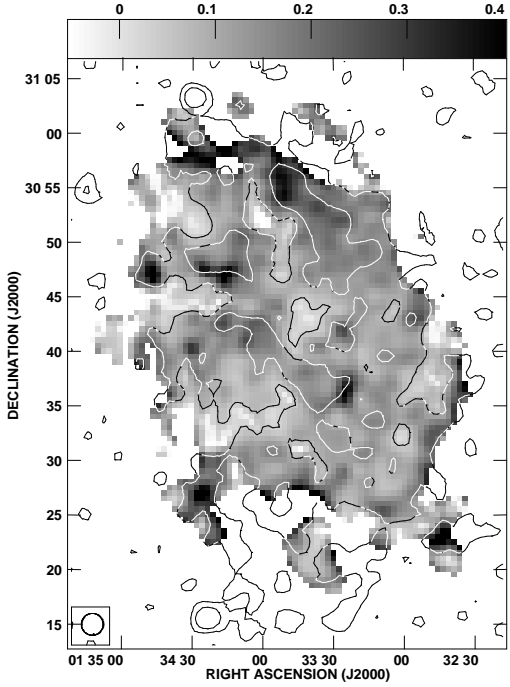}}
\caption[Nonthermal degree of polarization at 3.6\,cm]{Nonthermal degree of polarization at 3.6\,cm, with an angular resolution of 2${\arcmin}$ (the beam area is shown in the left-hand corner). Overlayed are contours of the linearly polarized intensity at 3.6\,cm with levels of 0.1 and 0.3 mJy/beam. }
\label{fig:pnth}
\end{center}
\end{figure}

Figure~\ref{fig:pnth} shows P$_{\rm nth}$ at 3.6\,cm. High nonthermal degrees of polarization (P$_{\rm nth}>$~30\%) are found in several patches of M33, with the most extended region of high P$_{\rm nth}$ in the northern part of the magnetic filament identified by \citet{Tabatabaei_2_07}, inside the second contour at DEC\,$>\,30^{\circ}\,54\arcmin$ in Fig.~\ref{fig:pnth}. The high P$_{\rm nth}$ also exist at 6.2\,cm in these regions.

Integrating the polarized and nonthermal intensity maps in the galactic plane out to galactocentric radius of R~$\leq$~7.5~kpc, we obtained the flux densities of the nonthermal ${\rm S}_{\rm nth}$ and linearly polarized ${\rm S}_{\rm PI}$ emission along with the average nonthermal degree of polarization $\bar{{\rm P}}_{\rm nth} = {\rm S}_{\rm PI}/{\rm S}_{\rm nth}$. Table~\ref{table:flux} gives ${\rm S}_{\rm nth}$, ${\rm S}_{\rm PI}$, and $\bar{{\rm P}}_{\rm nth}$ at different wavelengths, all at the same angular resolution of 180$\arcsec$.  At 3.6 and 6.2\,cm, $\bar{{\rm P}}_{\rm nth}$ is the same, demonstrating that Faraday depolarization effects are not significant at these wavelengths: the weaker S$_{\rm PI}$ at 3.6\,cm is due to lower synchrotron emissivity, as expected from the power-law behavior of synchrotron emission with respect to frequency. However, Faraday depolarization effects are possibly important at 20\,cm reducing $\bar{{\rm P}}_{\rm nth}$.

\section{Rotation measures}
\label{sec:RM}

When linearly polarized radio waves propagate in a magneto-ionic medium, their polarization vector is systematically rotated. The amount of Faraday rotation depends on the wavelength ($\lambda$) of the radio emission, the strength of the magnetic field along the line of sight (that is the regular field in the line of sight $\Brpa$), and the number density of thermal electrons ($n_e$) along the line of sight ($l$):
\begin{eqnarray}
\frac{\Delta\phi}{\rm rad} & = & 0.81 \left(\frac{\lambda}{\rm m}\right)^2  \int_0^{\frac{L}{\rm pc}} \left(\frac{\Brpa}{\mu{\rm G}}\right) \left(\frac{n_e}{{\rm cm}^{-3}}\right){\rm d}\left(\frac{l}{\rm pc}\right), \nonumber \\
 & = & 0.81\lambda^2\mathcal{R}
\label{eq:rotation}
\end{eqnarray}
where $L$ is the path length through the magneto-ionic medium.  Hence, the measured polarization angle ($\phi = \frac{1}{2}\, {\rm arctan}\frac{\rm U}{\rm Q}$) differs from the intrinsic polarization angle ($\phi_i$) as
\begin{equation}
\phi - \phi_i = \Delta\phi \equiv \lambda^2\,\mathcal{R}.
\label{eq:dphi}
\end{equation} 
When  Faraday depolarization is small (Faraday-thin condition), $\mathcal{R}$ does not depend on wavelength \citep{Sokoloff_98} and an estimate for $\mathcal{R}$ can be obtained from measurements of the polarization angles at two  wavelengths:
\begin{equation}
\frac{\RM}{\rm rad\,m^{-2}} = \frac{(\phi_1 / {\rm rad})\,-\, (\phi_2/{\rm rad})}{ \left(\frac{\lambda_1}{\rm m}\right)^2\, - \, \left(\frac{\lambda_2}{\rm m}\right)^2 }.
\label{eq:RM}
\end{equation} 
In this definition, the unknown intrinsic polarization angle of the source (or sources along the line of sight) cancels. Positive RM indicates that  $\Brpa$ points towards us.

Part of the measured RM is due to the interstellar medium of M33 (intrinsic RM or $\RMi$), the rest is due to the Galactic foreground medium ($\RMfg$), $\RM = \RMi + \RMfg$. The foreground rotation measure in the direction of M33 is mainly caused  by the extended Galactic magnetic bubble identified as region $A$ by \citet{Simard}. Assuming that the intrinsic contributions of the extragalactic sources 3C41, 3C42, and 3C48  (in a $5^{\circ}\times 5^{\circ}$ region around M33) themselves cancel out and the intragalactic contribution is negligible, \cite{Broten} and \cite{Tabara} found a foreground rotation measure of $-57\pm 10$\,rad~m$^{-2}$ for those sources. For the polarized sources in the $2^{\circ} \times2^{\circ} $ M33 field, \cite{Buczilowski_etal_91} found a foreground RM of $-55\pm 10$\,rad~m$^{-2}$. About the same value was obtained by \cite{Johnston}. In the following we use $\RMfg=-55\radm$. 

\begin{figure*}
\resizebox{\hsize}{!}{\includegraphics*{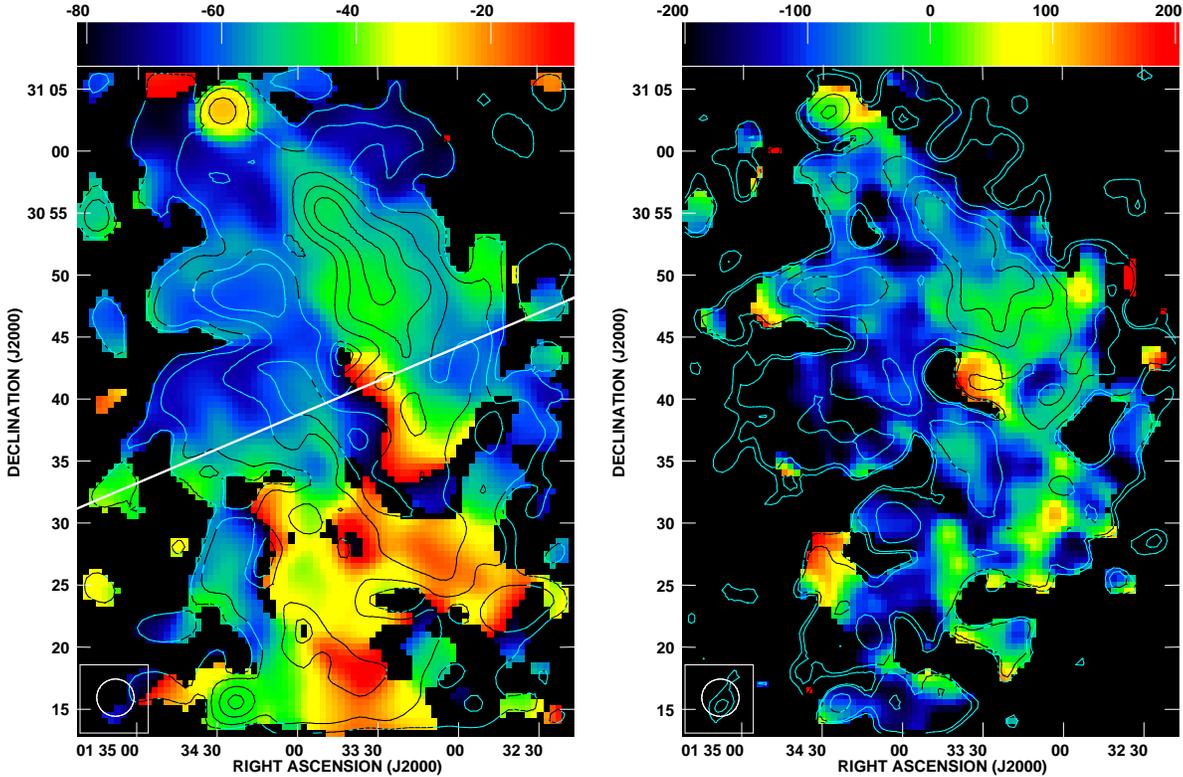}
\includegraphics*{RM3-6.color.ps}}
\caption{{\it Left:} observed rotation measure map of M33 ($\radm$) between 3.6 and 20\,cm with contours of 3.6\,cm polarized intensity. Contour levels are 0.1, 0.2, 0.4, 0.6, 0.8\,mJy/beam. {\it Right:} observed rotation measure between 3.6 and 6.2\,cm with contours of 6.2\,cm polarized intensity. Contour levels are 0.3, 0.4, 0.8, 1.2, 1.6\,mJy/beam. The angular resolution in both maps is 3${\arcmin}$ (the beam area is shown in the left-hand corners). The straight line in the left panel shows the minor axis of M33. }
\label{fig:rmtot}
\end{figure*}

Using the polarization data of \citet{Tabatabaei_2_07}, we first obtained the distribution of RM between  3.6 and 20\,cm across M33 (Fig.~\ref{fig:rmtot}, left panel), showing a smooth distribution of RM in the northern half of the galaxy. However, stronger and more abrupt fluctuations in RM occur in the southern half of the galaxy,  which are not due to the $\pm\, n\,73$\,rad\,m$^{-2}$ ambiguity in RM between these wavelengths ($\pm n \pi/ \mid \lambda_{1}^{2}-\lambda_{2}^{2} \mid $).
Weak polarized emission in the southern half at 20\,cm \citep[presented in ][]{Tabatabaei_2_07} can be linked to these RM variations. Between 3.6 and 6.2\,cm, RM varies less than between 3.6 and 20\,cm in the south of the galaxy (Fig.~\ref{fig:rmtot}, right panel).
This indicates that the relation between $\Delta\phi$ and $\lambda^{-2}$ in Eq.~(\ref{eq:dphi}) is not linear over the interval between 3.6 and 20\,cm due to Faraday depolarization at 20\,cm in the south of M33 and so RM measured at these wavelengths is not a good estimator for $\mathcal{R}$. 

\begin{figure*}
\begin{center}
\resizebox{7cm}{!}{\includegraphics*{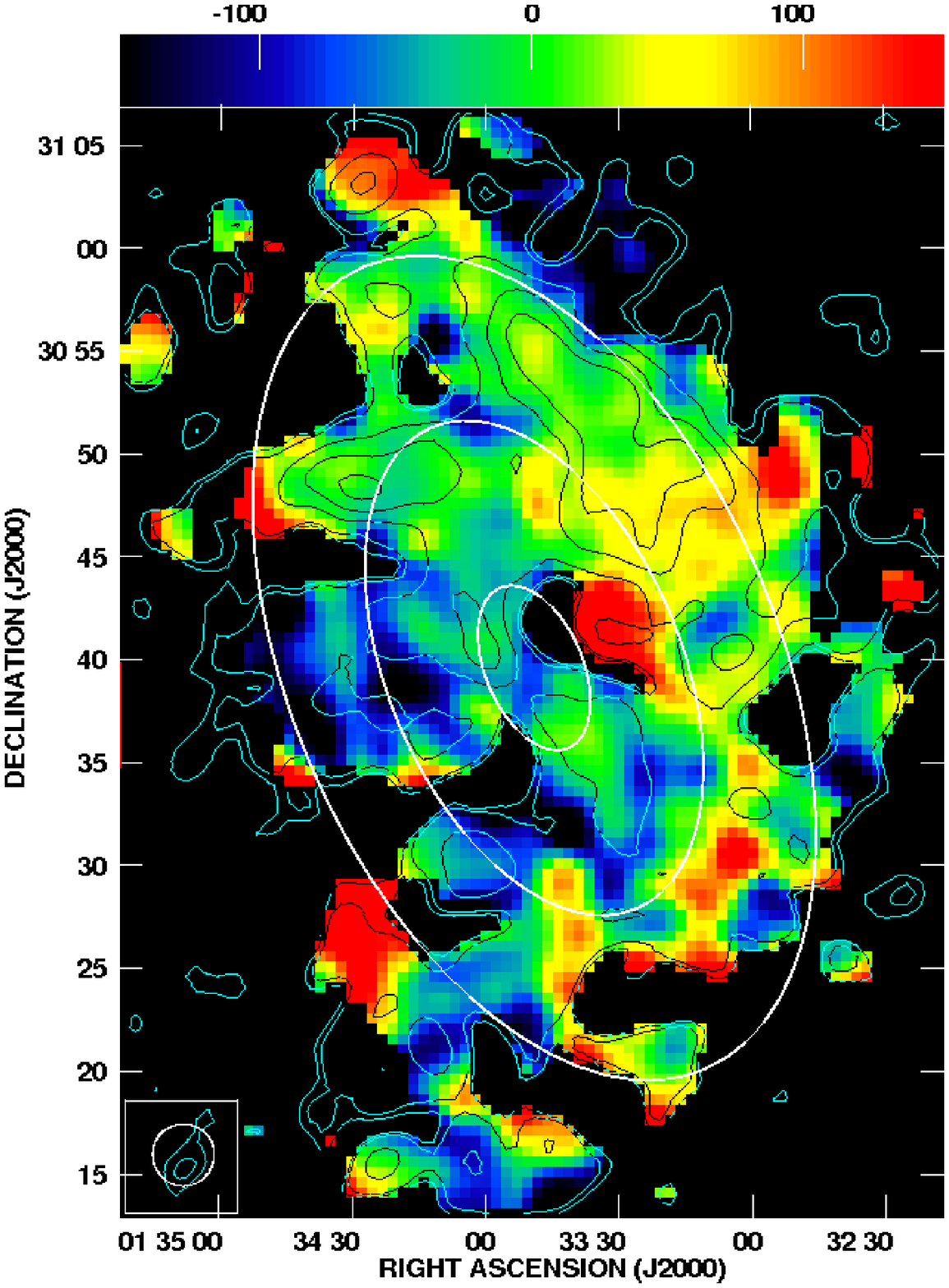}}
\resizebox{7cm}{!}{\includegraphics*{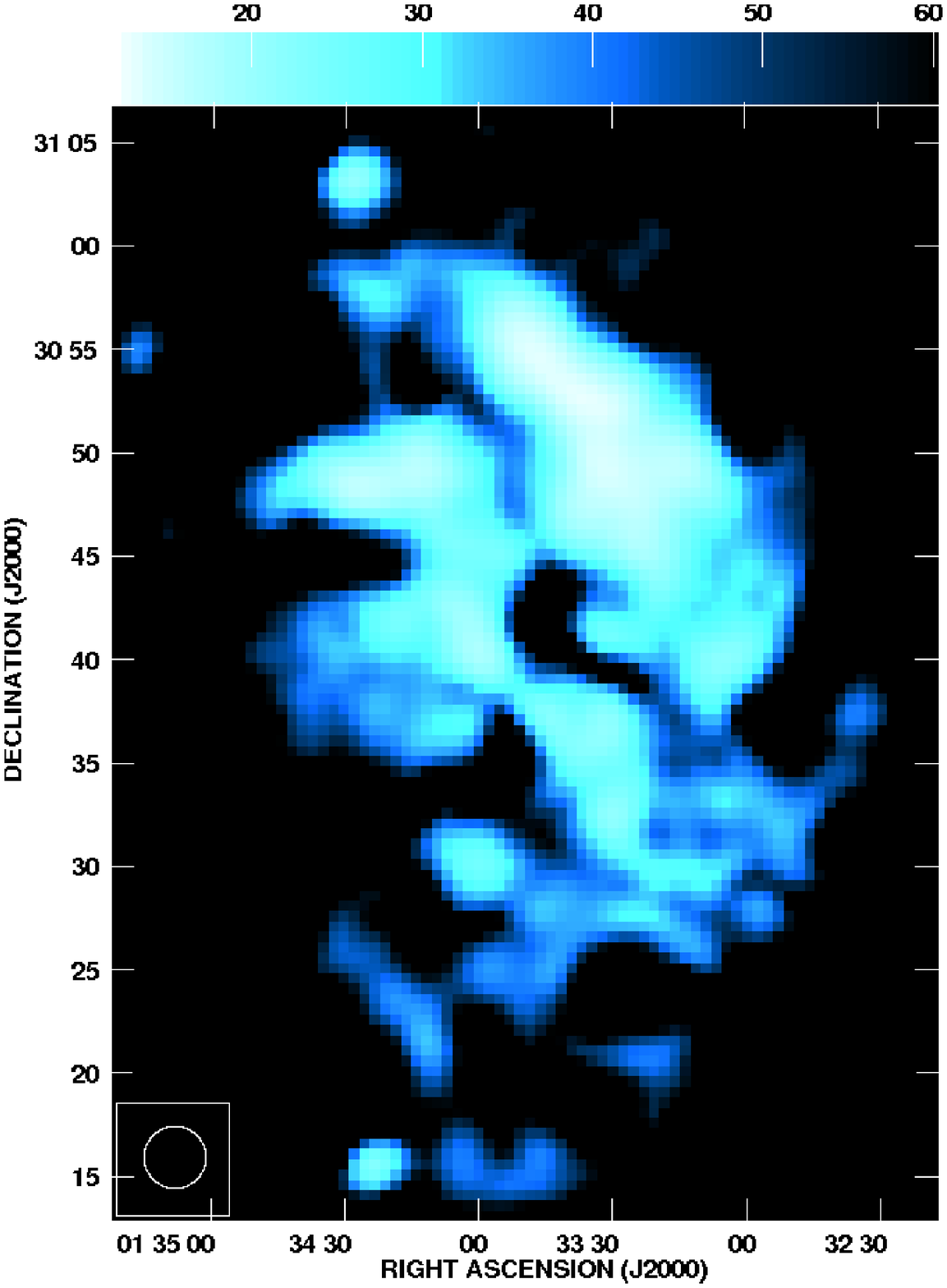}}
\caption{{\it Left:} rotation measure map of M33 ($\radm$) between 3.6 and 6.2\,cm after correction for the foreground $\RMfg=-55\radm$, with an angular resolution of 3${\arcmin}$ (the beam area is shown in the left-hand corner). Overlayed are contours of 6.2\,cm polarized intensity. Contour levels are 0.3, 0.4, 0.8, 1.2, 1.6, 3.2, 6.4\,mJy/beam. Also shown are rings at 1, 3, and 5\,kpc radii as used in Sect.~\ref{sec:mag}. {\it Right:} the distribution of the estimated error in RM is shown in the right panel.  }
\end{center}
\label{fig:rmint}
\end{figure*}

Figure~\ref{fig:rmint} shows $\RMi$ between 3.6 and 6.2\,cm which varies in a range including both positive and negative values. Comparing $\RMi$ with the overlayed contours of PI\footnote{Note that PI is related to the magnetic field in the plane of the sky that is a combination of both a mean field (or coherent regular field)  and anisotropic (compressed or sheared) random fields. $\RMi$ is related to only coherent regular field along the line of sight.}, $\RMi$ seems to vary more smoothly in regions of high PI. The apparent agreement between the ordered magnetic field in the plane of the sky and the regular magnetic field in the line of sight is  clearest in the north and along the minor axis and is best visible in the magnetic filament between the arms IV and V in the north-west of M33 \citep[Fig.~\ref{fig:Breg}, see also ][]{Tabatabaei_2_07} where $\RMi$ shows small variation within the PI contours. This indicates that the ordered magnetic field in this region is mainly regular.  
Sign variations of RM$_i$  are more frequent (arising locally) in the southern half, where there is no correlation with PI, than in the northern half of the galaxy. This indicates that the regular magnetic field is more affected by local phenomena \cite[like starforming activity, e.g. see ][]{Tabatabaei_2_07} in the south than in the north of the galaxy. 
The local RM$_i$ variations between large positive and negative values may represent loop-like magnetic field structures going up from and down to the plane (e.g. Parker loops). This is particularly seen in the central part of the galaxy besides regions in the southern arm II\,S (Fig.~\ref{fig:Breg}).


Figure~\ref{fig:rmint} also shows that the magnetic field is directed towards us on the western minor axis (at azimuth $\theta\approx 110\degr$ and $\theta\approx290\degr$), but has an opposite direction on the eastern side. The large RM$_i$ values in regions with small electron density, e.g. on the eastern and western minor axis and in a clumpy distribution in the central south near the major axis with 30$^{\circ}$~25$\arcmin<$DEC$<$30$^{\circ}$ 30$\arcmin$ 
\citep{Tabatabaei_3_07}, indicates a strong magnetic field along the line of sight and/or large path length through the magneto-ionic medium.  Considerable Faraday rotation measured on the eastern and western minor axis hints to deviation from a purely toroidal structure for the large-scale magnetic field \citep{Krause_90,Beck_96}. Furthermore, 
along with the magnetic field parallel to the disk, the presence of a vertical field component to the galactic disk is indicated in kpc-scale regions of large RM$_i$ but small $n_e$ values. The existence of the vertical magnetic field in M33 (that is strong near the major axis) is shown in Sect.~\ref{sec:mag}.


\section{The magnetic field}
\label{sec:mag}
\subsection{The regular magnetic field structure}

\begin{figure*}
\resizebox{\hsize}{!}{\includegraphics*{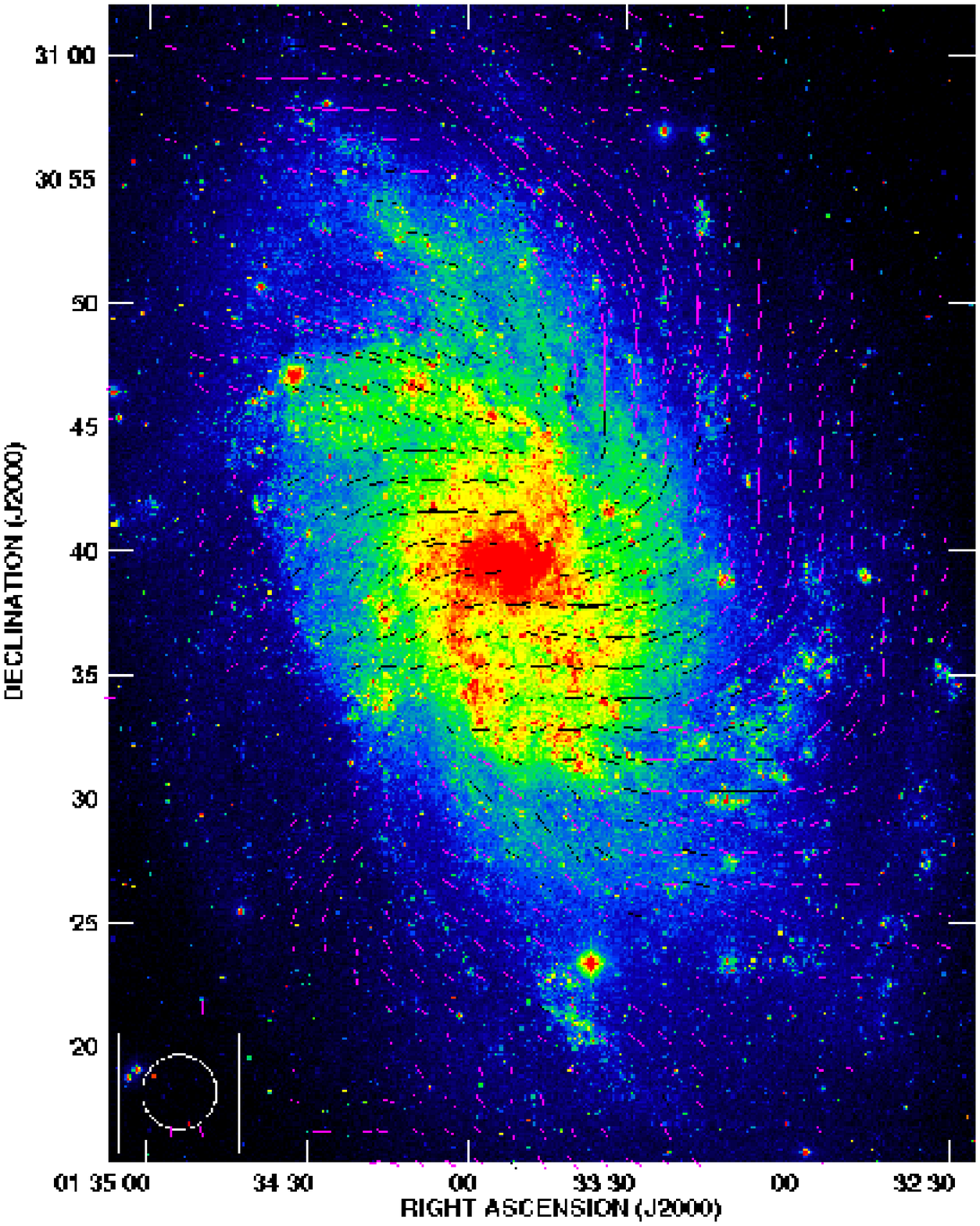}
\includegraphics*{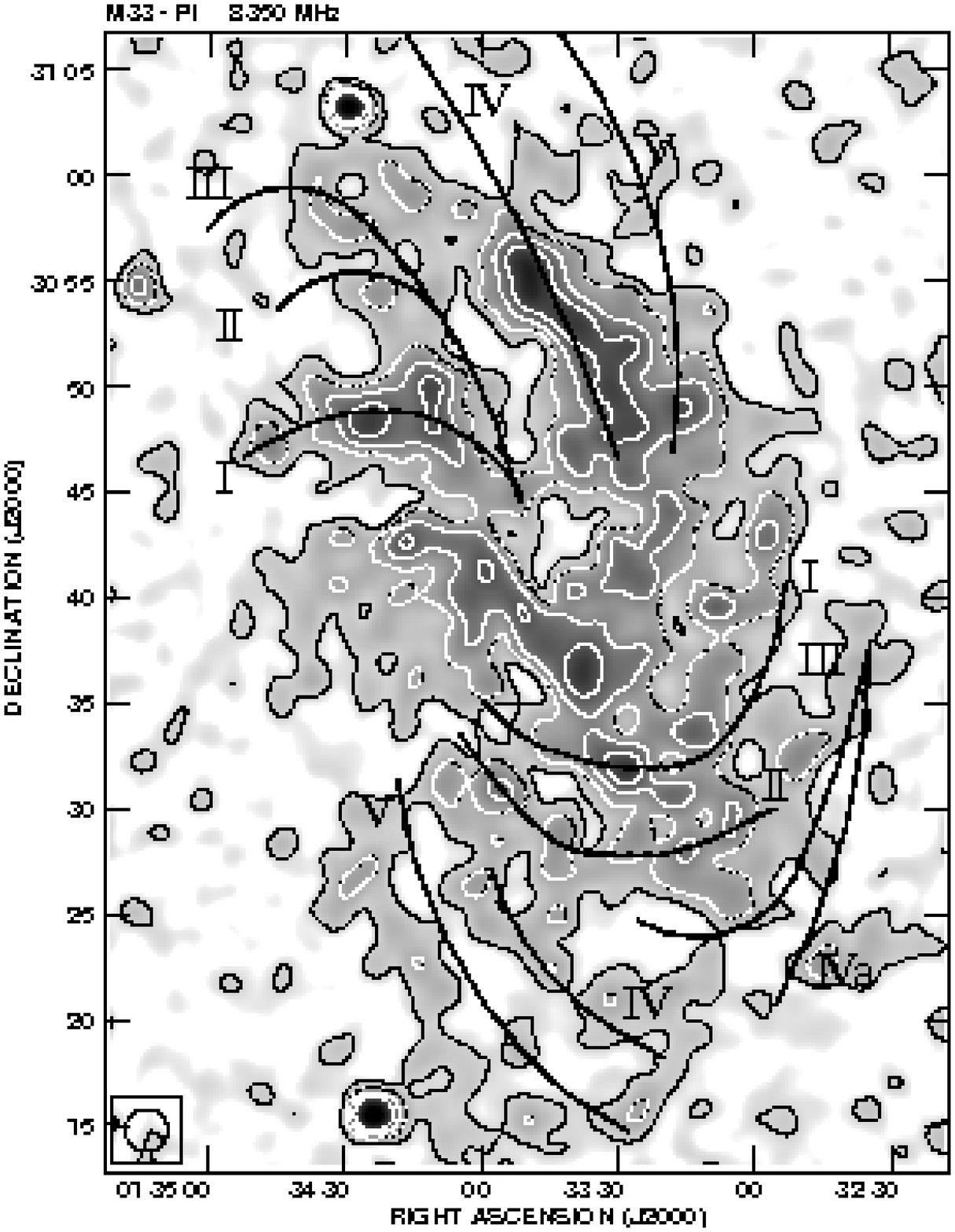}}
\caption{{\it Left:} Optical image (B-band, taken from the STScI Digitized Sky Survey) of M33 with overlayed vectors of intrinsic magnetic field in the sky-plane with 3$\arcmin$ angular resolution. The length gives the polarized intensity at 3.6\,cm, where 1$\arcmin$ is equivalent to 0.37\,mJy/beam. {\it Right:} A sketch of the optical arms \citep{Sandage} overlayed on the linearly polarized intensity with an angular resolution of 2$\arcmin$ (contours and grey scale) at 3.6\,cm \citep[see ][]{Tabatabaei_2_07}.  }
\label{fig:Breg}
\end{figure*}

The \wav{3.6} polarization angles were corrected for Faraday rotation and rotated by $90\degr$ to obtain the intrinsic orientation of the regular magnetic field component in the plane of the sky ($\Brpe$). Figure~\ref{fig:Breg} shows the derived $\Brpe$ superimposed on an optical image of M33. The orientation of $\Brpe$ shows a spiral magnetic field pattern  with a similar pitch angle to the optical arms in the north and south, but larger pitch angles in the east and west of the galaxy. 

The apparent continuity of the regular magnetic field straight through the center of M33 is remarkable. The Faraday rotation map (Fig.~\ref{fig:rmint}) shows that the field is oppositely directed on the east and west sides of the center. Unfortunately the current data lack the resolution to investigate the field properties in this region further.  

The behavior of $\Brpe$ along the minor axis ($\theta\approx 110\degr$, $\theta\approx290\degr$) is also informative. $\Brpe$ on the minor axis clearly has a strong radial component (Fig.~\ref{fig:Breg}). If we assume that $\Br$ lies solely in the disc, so that $\Brpa$ is produced by the galaxy's inclination (i.e. $\Buz=0$), then the change of sign in RM along the minor axis (Fig.~\ref{fig:rmint}) indicates that the direction of the radial component of $\vec{\Br}$ is towards center on both sides of the minor axis. This means that the dominant azimuthal mode of $\Br$ cannot be the bisymmetric $m=1$ mode suggested by earlier studies \citep{Beck_79, Buczilowski_etal_91}: we expect the dominant mode to be even. 

In order to identify the 3-D structure of the \emph{regular} magnetic field $\Br$ we fit a parameterized model of $\Br$ to the observed polarization angles at different wavelengths using the method successfully employed by \citet{Berkhuijsen_97} and \citet{Fletcher_04} to determine the magnetic field structures of M51 and M31 respectively. 

At each wavelength the maps in the Stokes parameters $Q$ and $U$ are averaged in sectors of $10\degr$ opening angle and $2\kpc$ radial width in the range $1\le r\le 5$ kpc. The size of the sectors is chosen so that the area of the smallest sector is roughly equivalent to one beam-area. We also take care that the standard deviation of the polarization angle in each sector is greater than the noise (if the sector sizes are too small then fluctuations in angle due to noise can be larger than the standard deviation). Then the
average $Q$ and $U$ intensities are combined to give the
average polarization angle $\psi=0.5\arctan{U/Q}$ and the average polarized emission intensity $\mathrm{PI}\sqrt{(Q^2+U^2)}$ in each sector. 

The observed polarization angles are related to the underlying properties of the regular magnetic field in M33 by\footnote{Note that we choose different symbols for the average polarization angle $\phi$ and the average polarization angle in sectors $\psi$.} 
\begin{equation}
  \label{eq:psi}
  \psi = \psi_\mathrm{0}(\Brpe) +
  \lambda^2\mathrm{RM_{i}}(\Brpa) + \lambda^2\RMfg,
\end{equation}
where $\psi_\mathrm{0}$ is the intrinsic angle of polarized emission,
$\mathrm{RM_{i}}$ is the Faraday rotation experienced by a photon as it passes through the magneto-ionic medium of M33 and $\RMfg$ is foreground Faraday rotation due to the Milky Way. $\Brpe$ is the component of the regular magnetic field of M33 that lies in the sky-plane and $\Brpa$ the component directed along the line-of-sight.

The cylindrical components of the regular field in the disk of M33 $\vec\Br=(\Bur,\But,\Buz)$ can be represented in terms of the Fourier series in the azimuthal angle $\theta$:
\begin{eqnarray}
  \label{eq:B}
  \Bur & = & B_0\sin p_0 + B_1\sin p_1\cos(\theta-\beta_1) \nonumber \\
      & & +\, B_2\sin p_2\cos2(\theta-\beta_2),\nonumber \\
  \But & = & B_0\cos p_0 + B_1\cos p_1\cos(\theta-\beta_1) \\
      & & +\, B_2\cos p_2\cos2(\theta-\beta_2),\nonumber\\
  \Buz & = & B_{z0} + B_{z1}\cos(\theta-\beta_{z1}) + B_{z2}
  \cos2(\theta-\beta_{z2}),\nonumber
\end{eqnarray}
where $B_m$ and $B_{zm}$ are the amplitude of the mode with azimuthal
wave number $m$ in the horizontal and vertical fields, $p_m$ is the
constant pitch angle of the $m$'th horizontal Fourier mode (i.e.\ the angle
between the field and the local circumference) and $\beta_m$ and
$\beta_{zm}$ are the azimuths where the non-axisymmetric modes are
maximum. The amplitudes of the Fourier modes are obtained in terms of the variables $B_m$ in units of $\FRM$ : in order to obtain amplitudes in Gauss independent information is required about the average thermal electron density and disc scale-height. Useful equations describing how $\Brpe$ and $\Brpa$ are related to the field components of Eq.~(\ref{eq:B}) can be found in Appendix A of \citet{Berkhuijsen_97}. 

Using Eqs.~(\ref{eq:psi}) and (\ref{eq:B}), and taking into account the inclination ($56\degr$) and major-axis orientation ($23\degr$) of M33, we fit the three-dimensional $\vec\Br$ to all of the observed polarization angles in a ring by minimizing the residual
\begin{equation}
  \label{eq:residual}
  S = \sum_{\lambda,n}\left\lbrack\frac{\psi_{n}
    - \psi(\theta_{n})}{\sigma_{n}}\right\rbrack^2 ,
\end{equation}
where $\psi_{n}$ is the observed angle of polarization,
$\psi(\theta_{n})$ the modelled angle in the sector centred on azimuth $\theta_{n}$ and $\sigma_{n}$ are the
observational errors. The Fisher test is used to ensure that the fits at each wavelength are statistically, equally good \citep[see Appendix B in][]{Berkhuijsen_97}. Errors in the fit parameters are estimated from the region of parameter space where $S\le \chi^2$ at the $2\sigma$ level.

\subsection{Results of fitting}
\label{subsec:fit}

\begin{figure}
\begin{center}
\includegraphics[width=0.39\textwidth]{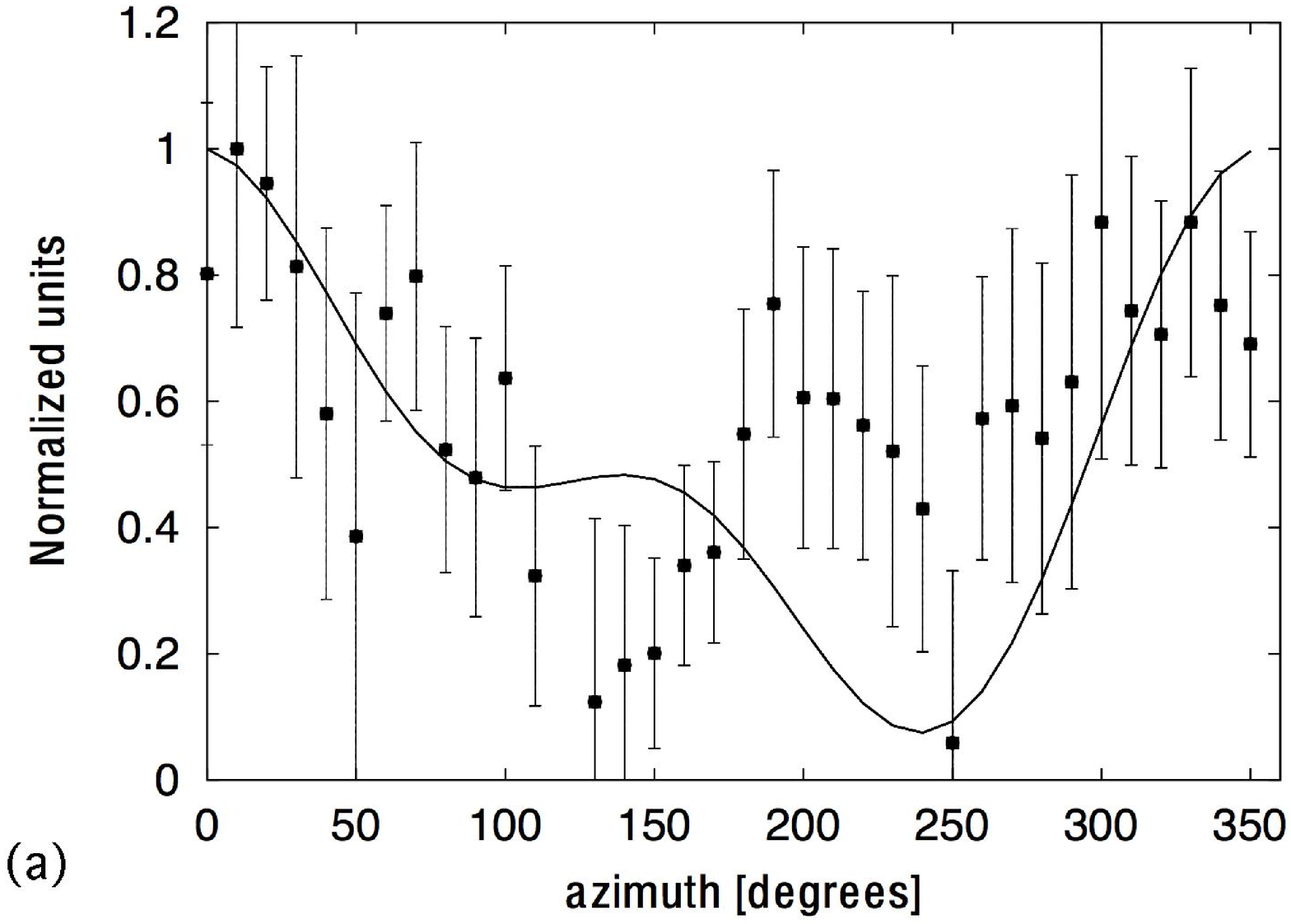}
\includegraphics[width=0.39\textwidth]{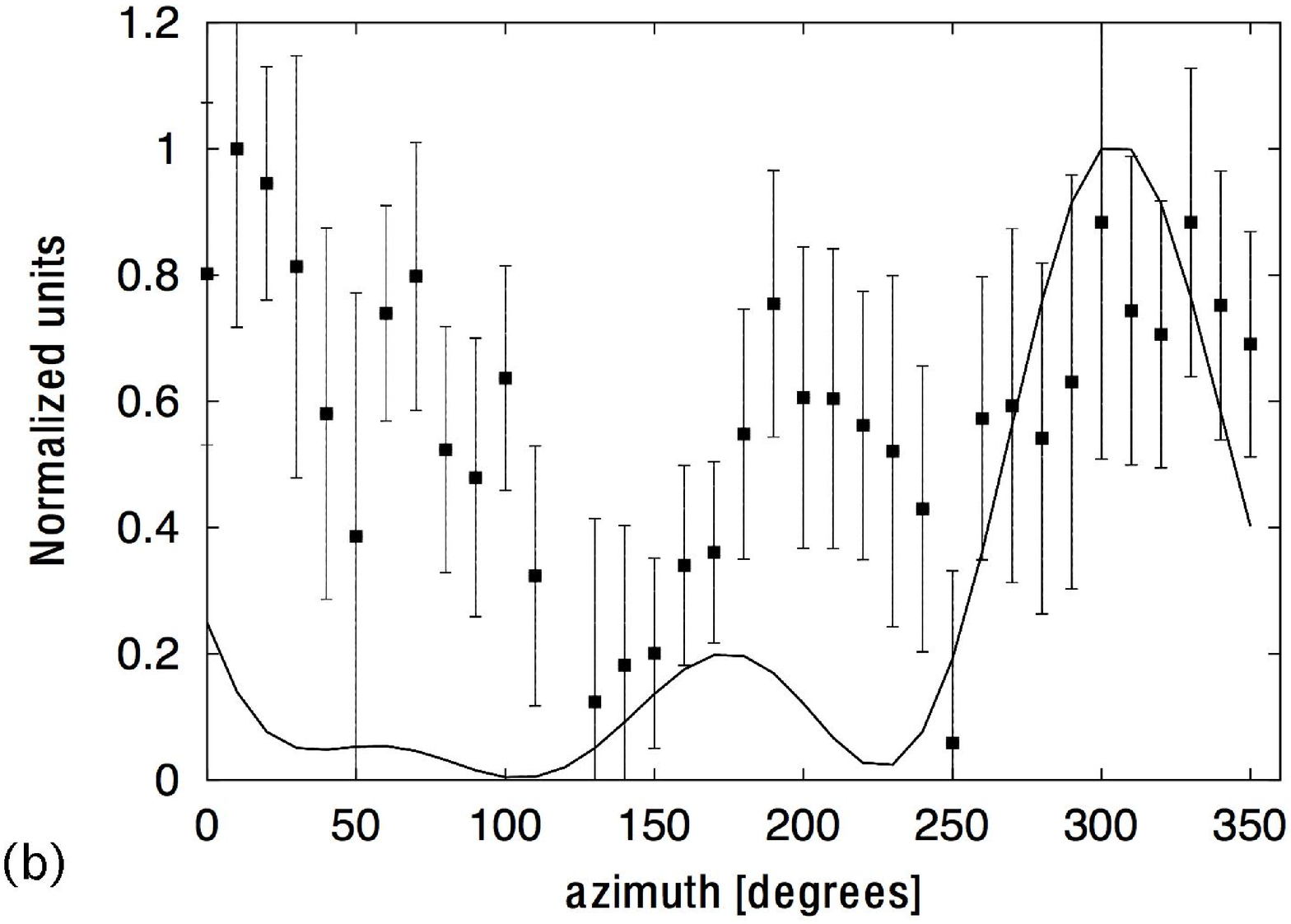}
\includegraphics[width=0.35\textwidth]{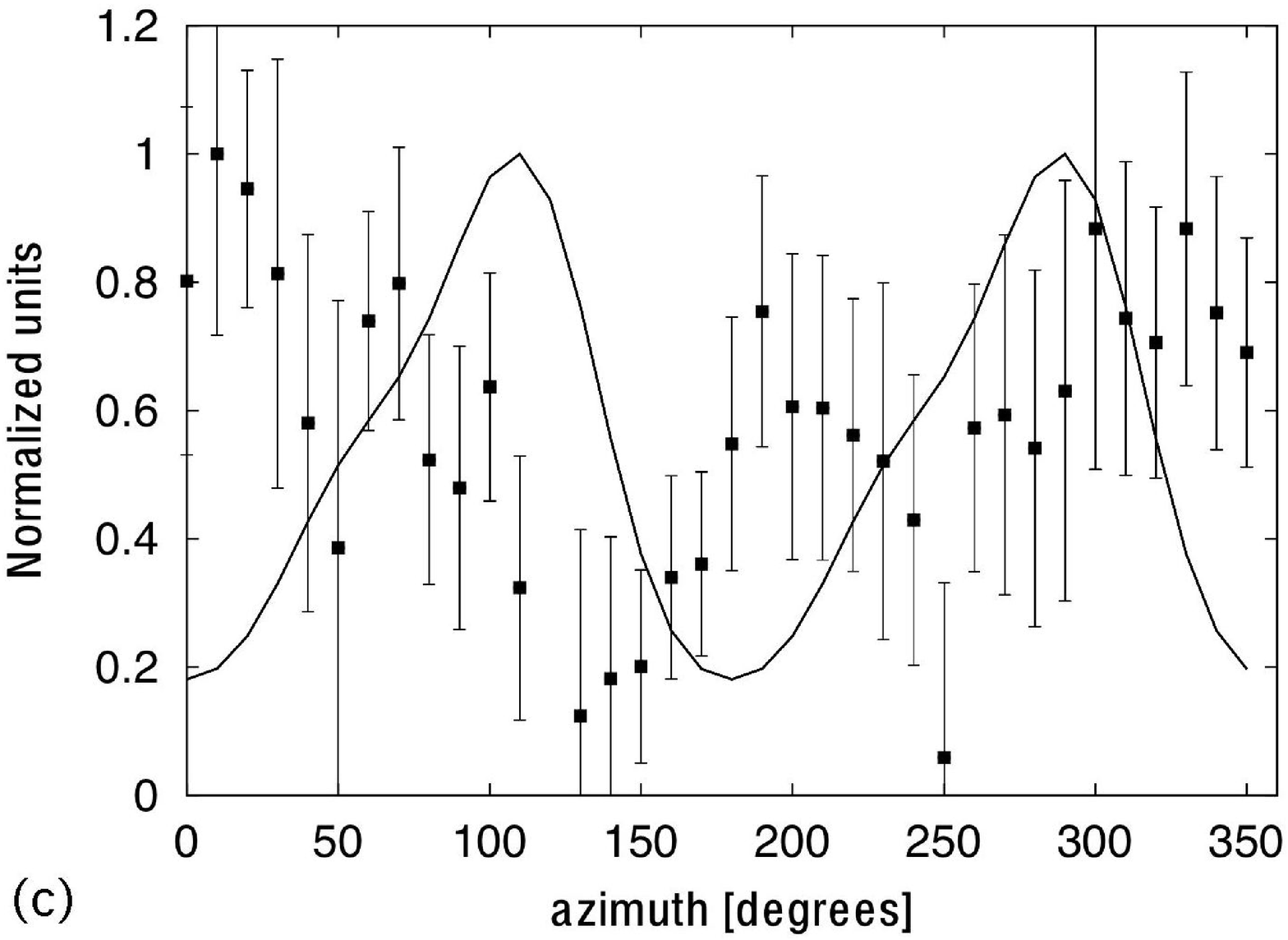}
\caption{Polarized intensity (PI) at 6\,cm in the ring 1--3 kpc, observed (points with error bars) and expected from the modelled regular magnetic field (lines) with PI$\propto\Brpe^2$ given by the different fits shown in Table.~\ref{tab:fit}. ({\it a})  using a fitted regular magnetic field containing the modes $m=0+z0+z1$. ({\it b}) as (a) but for the fit containing the modes $m=0+1+z1$. ({\it c}) as (a) but for the fit containing the modes $m=0+2$. }
\label{fig:model:pi}
\end{center}
\end{figure}

\begin{figure}
\begin{center}
\includegraphics[width=0.39\textwidth]{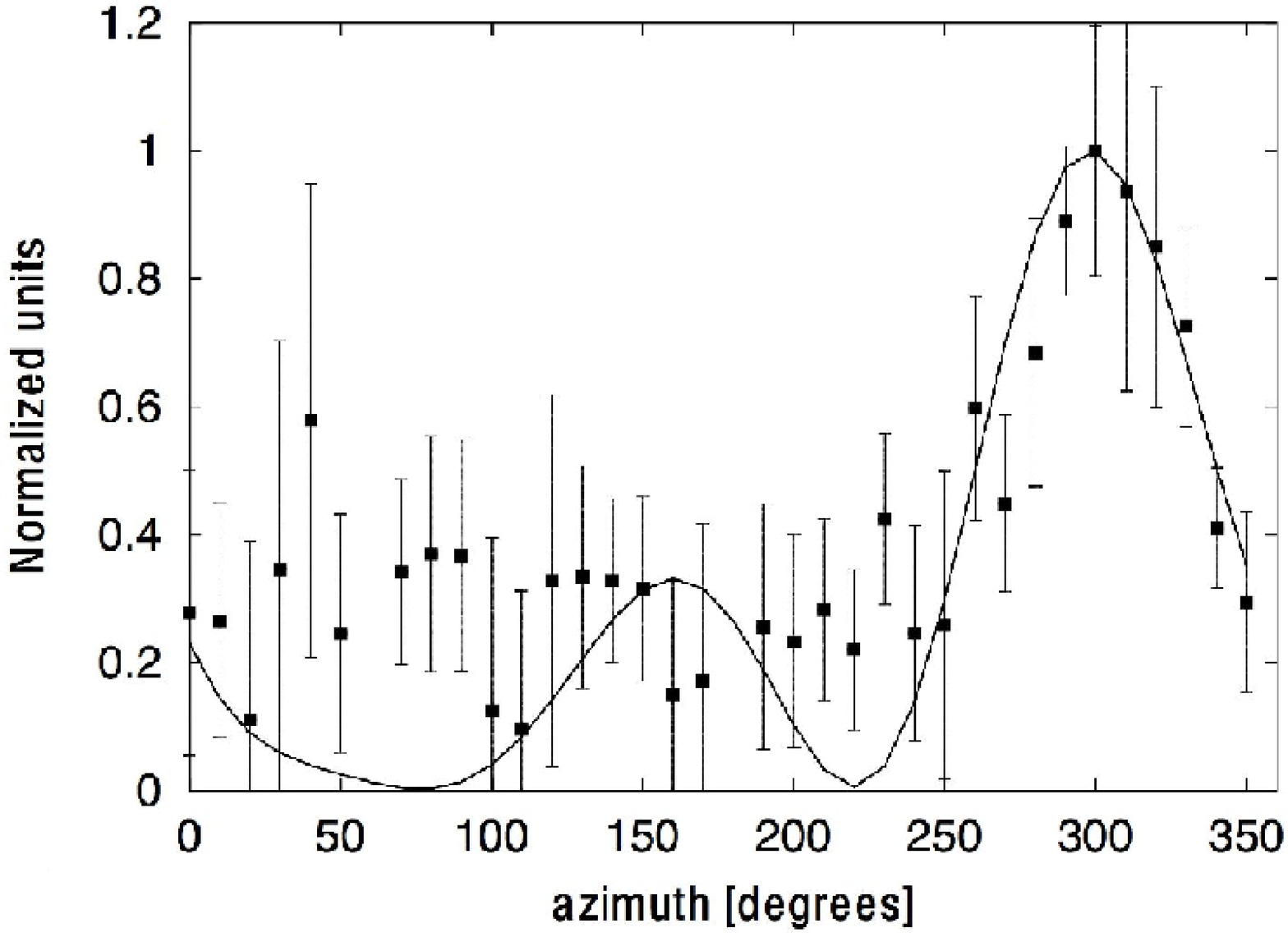}
\includegraphics[width=0.34\textwidth]{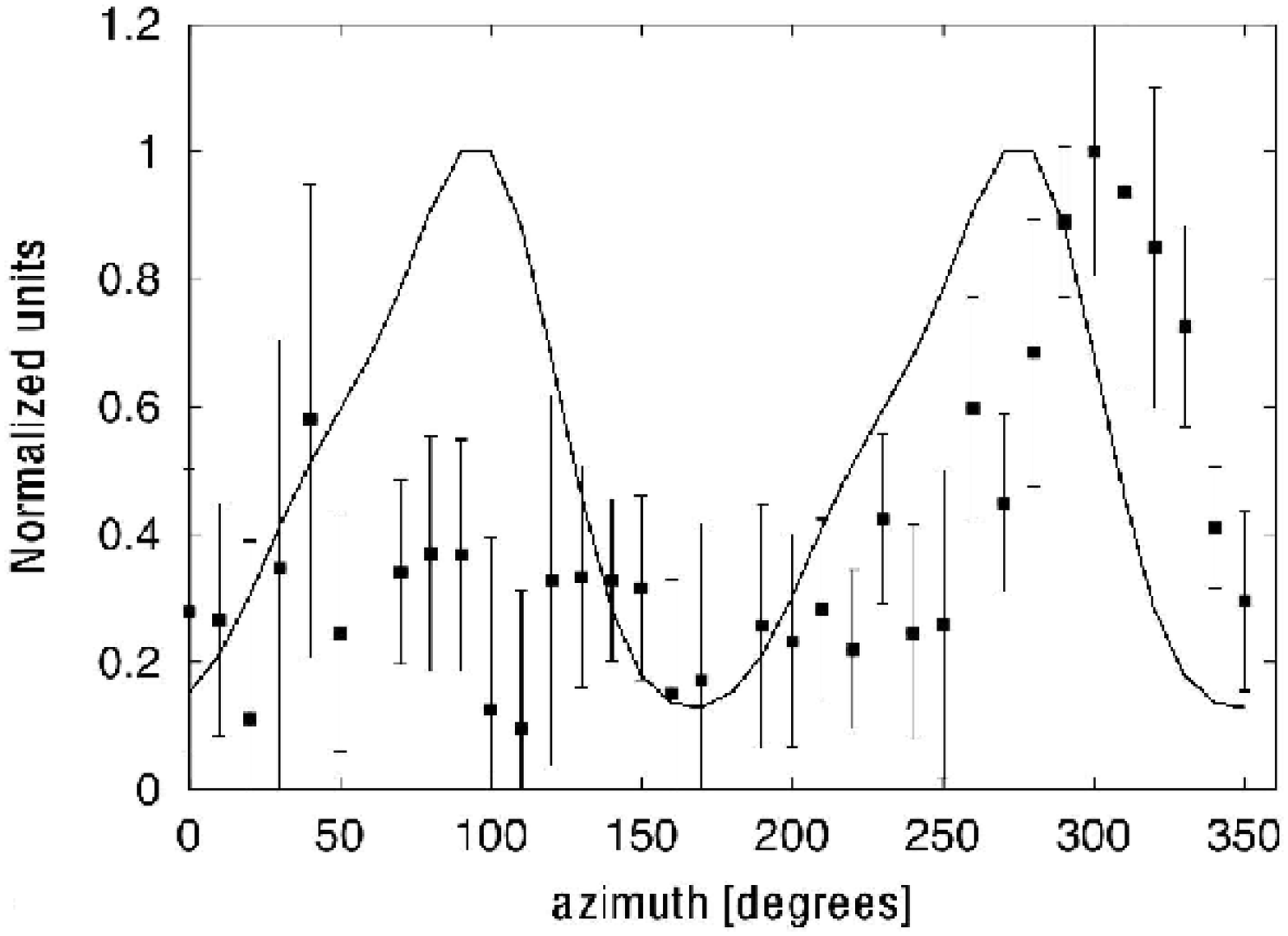}
\caption{As figure~\ref{fig:model:pi} but for the ring 3--5 kpc using a fitted regular magnetic field containing the modes $m=0+1+z1$ ({\it top})  and $m=0+2$ ({\it bottom}).}
\label{fig:model:pi2}
\end{center}
\end{figure}

We applied the method described in Sect.~\ref{sec:mag} to the polarization maps at 3.6 and 6.2\,cm. A preliminary examination of the data at all three wavelengths showed that the 20\,cm polarization angles do not have the same $\psi\propto\lambda^2$ Faraday rotation dependence as the angles at 3.6 and 6.2\,cm. This is most probably because the 20\,cm signal is strongly depolarized by Faraday effects (see Sect.~\ref{sec:depol}) and so only photons from an upper layer of the emitting region are detected in polarization. If the depolarization is constant at a given radius, a `depth' parameter can be used to take account of this effect \citep{Berkhuijsen_97, Fletcher_04}. However, in the case of M33 depolarization is strongly asymmetric (Fig.~\ref{fig:depol}) and this method does not lead to consistent results: we therefore work only with the polarization angles at 3.6 and 6.2\,cm to model the regular magnetic field. 

We fixed the foreground Faraday rotation, $\RMfg$ in Eq.~(\ref{eq:psi}), to $-55\FRM$ (see Sect.~\ref{sec:RM}). For each ring we found more than one statistically good fit to the observed polarization angles, three different fits in the ring 1--3 kpc and two fits in the ring 3--5 kpc, the fitted parameters are shown in Table~\ref{tab:fit}. All of the fits require the presence of more than one azimuthal Fourier mode and have two common characteristics: the presence of an $m=0$ mode that has a significant amplitude; the pitch angle of the $m=0$ mode is in the range $40\degr\lesssim p\lesssim 50\degr$. The reason why several equally good fits are found is the rather weak large-scale intrinsic rotation measure signal, i.e. the amplitude of systematic rotation measure variations is rather low compared to local fluctuations. This is a sign that the regular magnetic field of M33 is not as well-ordered and strong, relative to the small-scale field, as that of, for example, M31 \citep{Fletcher_04}. 

The model regular magnetic field given by Eq.~\ref{eq:B} is fitted to the observed polarization \emph{angles}, in order to obtain the results shown in Table~\ref{tab:fit}. Since we have not made use of the observed polarized \emph{intensity} we can try to use this to select the best regular magnetic field model for each of the two rings from the fits given in Table~\ref{tab:fit}. We compare the predicted azimuthal pattern of polarized intensity from the model fields with the observed polarized intensity (PI) at 6.2 (Faraday depolarization effects are negligible at 6.2\,cm and emissivity is higher than at 3.6\,cm due to the spectral index, thus giving a stronger signal). The model described by Eq.~\ref{eq:B} is not designed to reproduce the observed PI so we cannot make meaningful statistical assessments about the relative merits of the different fits in a given ring. But we can judge whether or not the fits are better or worse than each other in explaining the location of the main maxima and minima in the observed pattern of PI and so try to select a preferred model field for each ring.

Figures~\ref{fig:model:pi} and \ref{fig:model:pi2} shows the square of the predicted plane-of-sky regular magnetic field $\Brpe^2$ for each of the fits given in Table~\ref{tab:fit} and the observed 6.2\,cm polarized intensity, both normalized to avoid having to use a prescription for the poorly known synchrotron emissivity. In the case of energy equipartition between cosmic rays and magnetic fields the polarized intensity would be proportional to a higher power of $\Brpe$ than $2$ and maxima would be more pronounced. 

In the ring 1--3 kpc, the model field with the components of $m=0+z0+z1$ \footnote{ $z0$ and $z1$ are the first and secound Fourier modes of the vertical field.} (Fig.~\ref{fig:model:pi}a) reproduces the broad features of the observed polarized emission better than the models using the modes $m=0+1+z1$ (Fig.~\ref{fig:model:pi}b) and $m=0+2$ (Fig.~\ref{fig:model:pi}c). The match to the observed PI is far from perfect in Fig.~\ref{fig:model:pi}a but this is the only model field that can account for the strong excess of PI at 6\,cm in the northern half of the disc at these radii. In the ring 3--5 kpc the fit using $m=0+1+z1$ (Fig.~\ref{fig:model:pi2}, top) is better at reproducing the general pattern of polarized emission at 6\,cm than the other statistically good fit using $m=0+2$ (Fig.~\ref{fig:model:pi2}, bottom). Again, the match to observations is not perfect, but the model with $m=0+2$ would produce a strong maximum in PI at $\theta\sim 90\degr$ that is not observed. 

\begin{figure}
\begin{center}
\includegraphics[width=0.5\textwidth]{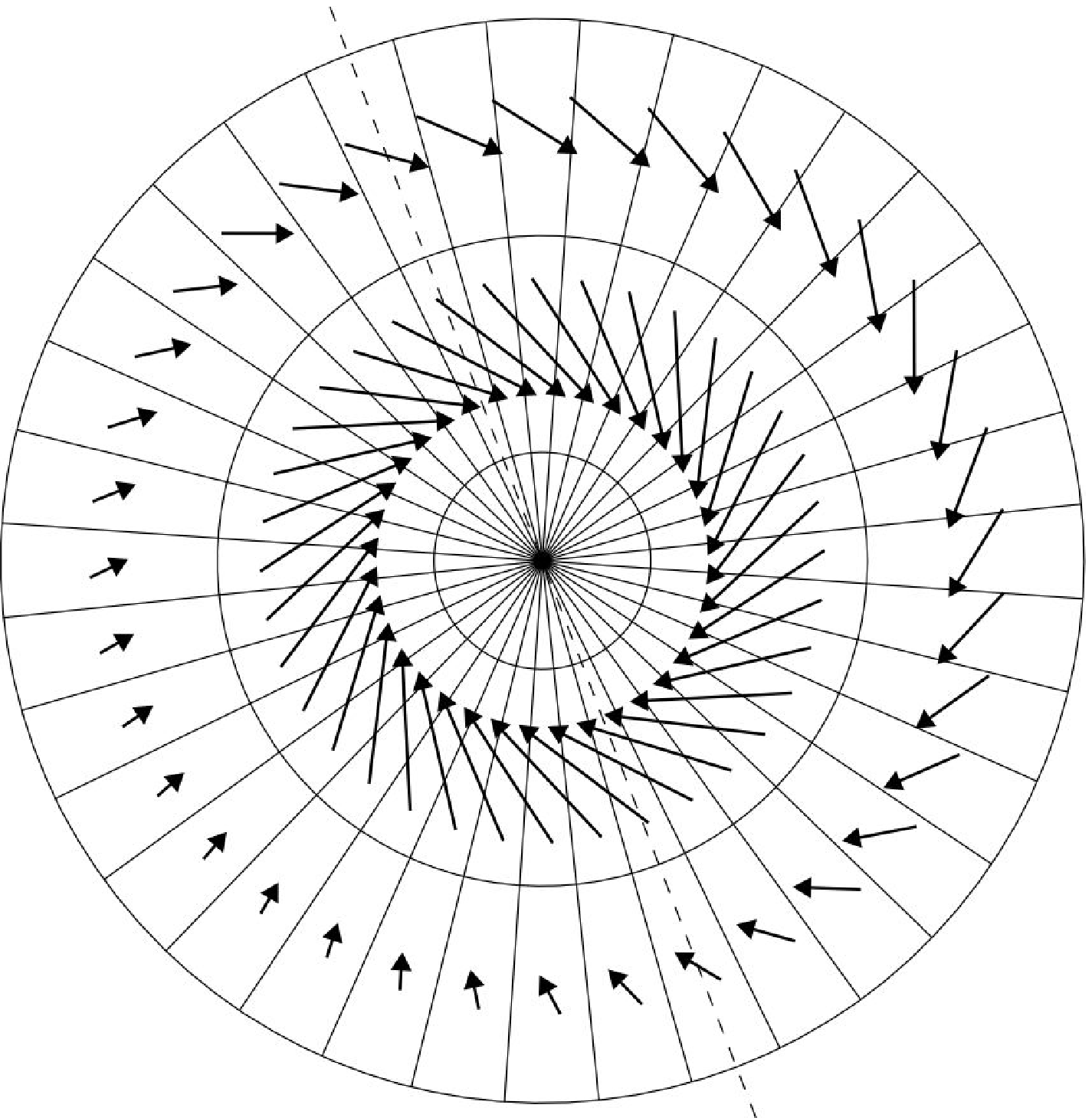}
\caption{The regular magnetic field described by our favoured fitted model (see Sect.~\ref{subsec:fit} for details). The galaxy has been deprojected into a face-on view, so only the disk-plane components of the magnetic field are shown, the vertical components are not visible. The dashed line shows the major axis of M33, with north to the top. The ring boundaries are at 1, 3 and 5~ kpc.}
\label{fig:polar}
\end{center}
\end{figure}

To summarize: we select the statistically good fit using the modes $m=0+z0+z1$ in the ring 1--3 kpc and that using $m=0+1+z1$ in the ring 3--5 kpc as being the best descriptions of the regular magnetic field in M33 (the parameters of these two preferred fits are given in columns 3 and 6 of Table~\ref{tab:fit}). Our reason is that the $\Brpe^2$ produced  by these models produces a much closer match to the observed pattern of PI at 6.2\,cm than other statistically good models. Fig.~\ref{fig:polar} shows the regular magnetic field in a face-on view of the galaxy (thus the vertical field components cannot be seen).

\begin{table*}
\caption{Parameters of the fitted models and their $2\sigma$ errors.
$RM_\mathrm{fg}$ is the Faraday rotation measure arising in the Milky Way,
$B_m$ and $p_m$ are the amplitude and pitch angle of the mode with wave number
$m$, and $\beta_m$ is the azimuth where a mode with azimuthal wave number $m$
is maximum. The minimum value of the residual $S$ and the value of $\chi^2$ are
shown  for each model in the bottom lines.  The combination of azimuthal modes of $m=0+z0+z1$ and $m=0+1+z1$ can best re-produce the observed polarized intensity in the 1-3 and 3-5\,kpc rings, respectively (Figs~\ref{fig:model:pi} and \ref{fig:model:pi2}). }
\begin{center}
\begin{tabular}{llcccccc} \hline \noalign{\smallskip}
  & Units & \multicolumn{6}{c}{Radial range (kpc)}  \\
  & & \multicolumn{3}{c}{1--3} & & \multicolumn{2}{c}{3--5} \\
\hline \noalign{\medskip} 
RM$_{\rm fg}$ & $\rm{rad\,m^{-2}}$ &
  $-55$\,\scriptsize{$\pm45$} & 
  $-55$\,\scriptsize{$^{+6}_{-9}$} &
  $-55$\,\scriptsize{$^{+30}_{-59}$} & 
  &
  $-55$\,\scriptsize{$^{+30}_{-60}$} &
  $-55$\,\scriptsize{$\pm19$}  \\ \noalign{\smallskip}
$B_0$ & $\rm{rad\,m^{-2}}$ &
  $-30$\,\scriptsize{$^{+11}_{-20}$} &
  $-69$\,\scriptsize{$\pm4$} & 
  $-14$\,\scriptsize{$\pm2$} & 
  &
  $-13$\,\scriptsize{$\pm3$} &
  $-103$\,\scriptsize{$\pm9$}  \\ \noalign{\smallskip}
$p_0$ & deg &
  $48$\,\scriptsize{$\pm12$} & 
  $51$\,\scriptsize{$\pm2$} & 
  $42$\,\scriptsize{$\pm4$} & 
  &
  $42$\,\scriptsize{$^{+1}_{-7}$} & 
  $41$\,\scriptsize{$\pm2$}  \\ \noalign{\smallskip}
$B_1$ &  $\rm{rad\,m^{-2}}$ &
    & 
    &
    $-12$\,\scriptsize{$\pm3$}&
    &
    $-9$\,\scriptsize{$\pm2$} &
    \\ \noalign{\smallskip}
$p_1$ & deg &
    & 
    &
    $28$\,\scriptsize{$^{+1}_{-9}$} &
    &
    $14$\,\scriptsize{$^{+11}_{-7}$} &
    \\ \noalign{\smallskip}
$\beta_1$ & deg &
    &
    &
    $-56$\,\scriptsize{$^{+12}_{-1}$} &
    &
    $-67$\,\scriptsize{$^{+22}_{-39}$} &
    \\ \noalign{\medskip}
$B_2$ &  $\rm{rad\,m^{-2}}$ &
    & 
    $-41$\,\scriptsize{$\pm3$} &
    &
    &
    &
    $-67$\,\scriptsize{$\pm6$} \\ \noalign{\smallskip}
$p_2$ & deg &
    & 
    $-87$\,\scriptsize{$\pm6$}&
    &
    &
    &
    $-103$\,\scriptsize{$\pm8$}\\ \noalign{\smallskip}
$\beta_2$ & deg &
    &
    $-12$\,\scriptsize{$\pm3$}&
    &
    &
    &
    $-22$\,\scriptsize{$\pm4$}\\ \noalign{\medskip}
$B_{z0}$ & $\rm{rad\,m^{-2}}$ &
   $-14$\,\scriptsize{$\pm15$} &
   &
   &
   &
   & 
   \\ \noalign{\smallskip}
$B_{z1}$ &  $\rm{rad\,m^{-2}}$ &
  $-52$\,\scriptsize{$^{+22}_{-35}$} &
  &
  $-16$\,\scriptsize{$\pm5$} &
  &
  $-14$\,\scriptsize{$\pm4$} &
  \\ \noalign{\smallskip}
$\beta_{z1}$ & deg &
  $32$\,\scriptsize{$^{+21}_{-16}$} &
  &
  $24$\,\scriptsize{$^{+1}_{-10}$} &
  &
  $9$\,\scriptsize{$\pm9$} &
  \\ \noalign{\smallskip}
\hline \noalign{\medskip}
$m$  &  & $0+z0+z1$ & $0+2$ & $0+1+z1$ & & $0+1+z1$ & $0+2$ \\ \noalign{\smallskip}
$S$ & & $56$ & $83$ & $57$ & & $57$ & $82$ \\ \noalign{\smallskip}
$\chi^2$ & & $85$ & $85$ & $84$ & & $84$ & $85$ \\
\noalign{\medskip}
  \hline
  \label{tab:fit}
\end{tabular}
\end{center}
\end{table*}

\subsection{The equipartition magnetic field strengths}
\label{subsec:equipart}

The strengths of the total magnetic field $\Bt$ and its regular component $\Br$ can be found from the total synchrotron intensity and its degree of linear polarization P$_{\rm nth}$. Assuming equipartition between the energy densities of the magnetic field and cosmic rays ($\varepsilon_{CR} = \varepsilon_{\Bt} = \Bt^2/8\pi$), 

\begin{eqnarray}
\Bt= \big[\frac{ 4\pi (2\alpha_n+1)\, {\mathrm{K}}' \,I_{n} \,
E_\mathrm{p}^{1-2\alpha_n} \, (\frac{\nu}{2 c_1})^{\alpha_n}} { (2\alpha_n-1)\, c_2(\alpha_n)\, L\, c_3}\big]^{\frac{1}{\alpha_n+3}}
\label{eq:Btoteq}
\end{eqnarray}
\citep{Beck_06}, where ${\mathrm{K}}'={\mathrm{K}} +1$  with ${\mathrm{K}}$ the ratio between the number densities of cosmic ray protons and electrons, $I_n$ is the nonthermal intensity,  $L$ the pathlength through the synchrotron emitting medium, and $\alpha_n$ the mean synchrotron spectral index. $E_\mathrm{p} = 938.28$\,MeV\,$=1.50\times 10^{-3}$\,erg is the proton rest energy and
\noindent
\begin{eqnarray}
c_1 & = & \left. 3e/(4\pi {m_\mathrm{e}}^3 c^5) = \frac{6.26428\cdot 10^{18}}{{\rm erg}^{2}.\,{\rm s.\, G}}, \right. \nonumber \\
c_2(\alpha_n) & = & {1\over4} c_3\,\left(\alpha_n+5/3\right)/(\alpha_n+1) \, \Gamma [(3\alpha_n+1)/6] \nonumber \\
& & \times \, \Gamma [(3\alpha_n+5)/6]. \nonumber 
\end{eqnarray}
For a region where the field is completely regular and has a constant inclination $i$ with
respect to the sky plane ($i=0^o$ is the face-on view), $c_3 = [{\cos}\,(i)]^{(\alpha_n+1)}$. If the field is completely turbulent and has an isotropic angle distribution in three dimensions, $c_3 = (2/3)^{(\alpha_n+1)/2}$.
If the synchrotron intensity is averaged over a large volume, $[{\cos}\,(i)]^{(\alpha_n+1)}$ has to be replaced by its average over all occurring values of $i$.

The strength of the regular magnetic field in the plane of the sky can be estimated from the observed nonthermal degree of polarization \citep{Segalovitz}:

\begin{eqnarray}
{\rm P}_{\rm nth} & = & \left. \left(\frac{3\gamma\,+\,3}{3\gamma\,+\,7}\right)\, \times  \right. \nonumber \\ 
& & \left. \,\,\left[1 + \frac{(1-q)\,\pi^{1/2}\,\Gamma[(\gamma + 5)/4]} {2q \Gamma[(\gamma + 7)/4]F(i)}\right]^{-1},\right.\nonumber \\ 
F(i) & = & \frac{1}{2\pi}\,\int_0^{2\pi}\, \left(1-{\sin^2{i}}\,\sin^2{\theta}\right)^{(\gamma +1)/4}\,{\rm d}\theta, \nonumber \\
\label{eq:Bregeq}
\end{eqnarray}
with $\Br/\bt = q^{2/(1+\gamma)}$, $\gamma = 2\alpha_n +1$, and $\theta$ the azimuthal angle ($\bt$ is the turbulent magnetic field). This formula assumes that the regular magnetic field has a single orientation, is parallel to the disk and, taken over the galaxy as a whole, has no further preferential orientation with respect to any fixed direction in space.

The determined average values of $I_n$, $\alpha_n$, and P$_{\rm nth}$ with the assumed values of ${\mathrm{K}}\,(\simeq$\,100) and  $L\,(\simeq\,1\,{\rm kpc}/ {\rm cos}\,i$) lead to  $\Bt =\, 6.4\,\pm \,0.5\,\mu$G  and $\Br =\, 2.5\,\pm\,1.0\,\mu$G for the disk of M33 ($R<$7.5\,kpc). The strongest regular magnetic field is found in between the northern arms IV\,N and V\,N (in the magnetic filament) with $\Br\simeq 6.6\,\mu$G where $\Bt \simeq 8.3\,\mu$G. 

The regular magnetic field strength estimated from the mean rotation measure in sectors and assuming $n_e\simeq 0.03$\,cm$^{-3}$ (see Sect.~\ref{sec:depol}) is $1.4\,\scriptsize{ ^{+0.9}_{-0.5}}$  $\mu$G and $0.6\pm 0.1\,\mu$G in the ring 1--3\,kpc and 3--5\,kpc, respectively.  In the ring 1--3\,kpc, the regular field strength is consistent with that estimated from the equipartition assumption, while it is much smaller in the ring 3--5\,kpc.  The more frequent RM variations in both amplitude and sign in the second ring (see Fig.~3) indicate a line-of-sight field component with many reversals and a small mean RM within each sector, whereas the field component in the sky plane causes significant polarized emission (which is insensitive to the field reversals).

The equipartition total magnetic field strength of $6\,\pm\,0.5\mu$G is slightly higher than the total field obtained by \cite{Buczilowski_etal_91}, $4\,\pm\,1 \mu$G, assuming the minimum total energy requirement of the disk-like synchrotron source. This is not surprising because a difference of $\simeq$20\% is expected between the minimum and equipartition magnetic field strengths for field strengths of about $ 5\,\mu$G \citep{Beck_06}. Furthermore, the mean nonthermal fraction from the standard thermal/nonthermal separation method is lower than that from the new method \citep{Tabatabaei_3_07}, resulting in a weaker equipartition magnetic field. 

\begin{figure}
\begin{center}
\resizebox{8cm}{!}{\includegraphics*{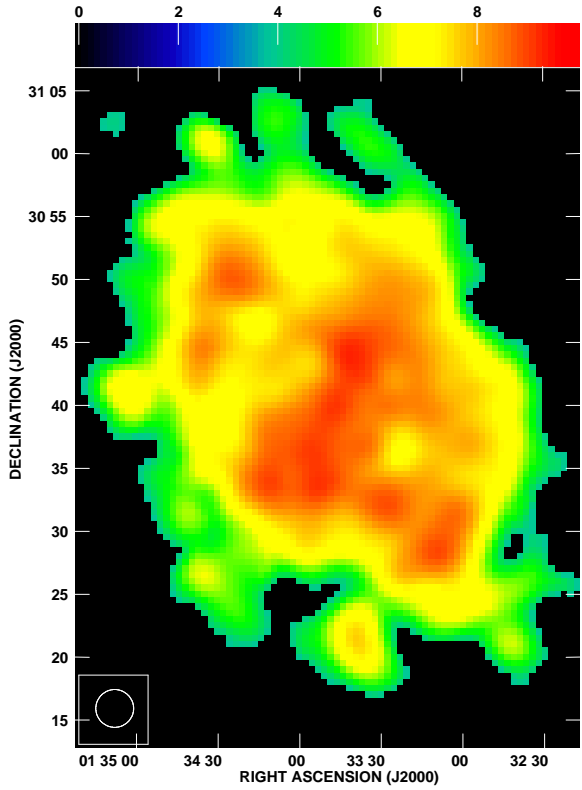}}
\caption{Distribution of the turbulent magnetic field strength, $\bt$ ($\mu$G), in M33 with an angular resolution of 3${\arcmin}$ (the beam area is shown in the left-hand corner).    }
\label{fig:Br}
\end{center}
\end{figure}

As the polarized intensity (PI/0.75, corrected for the maximum fractional polarization in a completely regular field) is related to the regular magnetic field strength, and the nonthermal intensity ($I_{n}$) to the total magnetic field strength in the plane of the sky, $I_n\,-\,$(PI/0.75) gives the nonthermal emission due to the turbulent magnetic field $\bt$. Using this intensity with Eq.~(\ref{eq:Btoteq}) and assuming a completely turbulent field yields the distribution of $\bt$ across the galaxy. Figure~\ref{fig:Br} shows strong  $\bt$ ($>7\,\mu$G) in the central region of the galaxy, the arm I\,S, and parts of the northern arm I\,N.   

\begin{figure}
\begin{center} 
\resizebox{8cm}{!}{\includegraphics*{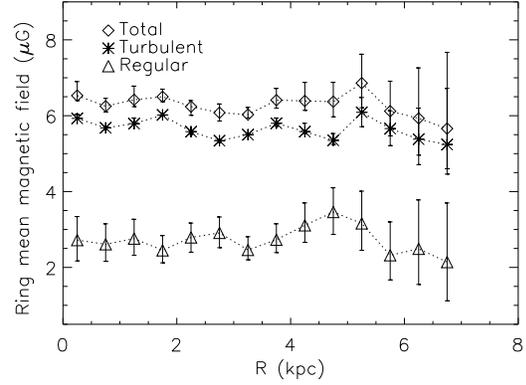}}
\caption{Variation of the mean total, regular, and turbulent magnetic field strengths in rings of 500\,pc width with galactocentric radius.} 
\label{fig:bring}
\end{center}
\end{figure}

Using the mean synchrotron flux density, synchrotron spectral index, and degree of polarization in rings, we also derive the average field strengths in rings.  Figure~\ref{fig:bring} shows some fluctuations but no systematic increase or decrease of these strengths with galactocentric radius. The small bump at $4.5<R<5.5$\,kpc is due to the M33's magnetic filament.

\section{Depolarization}
\label{sec:depol}

The depolarization observed at a certain wavelength is defined as the ratio of the nonthermal degree of linear polarization P$_{\rm nth}$ and the theoretical maximum value $p_0$ (75\% for $\alpha_n=\,1$). Generally, depolarization may be caused by instrumental effects as the bandwidth  and beamwidth of the observations or by the wavelength-dependent Faraday depolarization. 
Bandwidth depolarization occurs when the polarization angles vary across the frequency band, reducing the observed amount of polarized emission. It is given by $sinc\,(2{\rm RM}\,{\lambda}^2\, \delta \nu/\nu)$, where $\delta \nu$ is the bandwidth of the observations \citep[e.g. ][]{Reich_06}. In our study, the wavelengths, bandwidths and RM values lead to a negligible bandwidth depolarization.
Beamwidth depolarization occurs when polarization vectors of different orientation are unresolved in the telescope beam. In order to compensate this effect, the ratio of the nonthermal degree of polarization  at two wavelengths is used at a same angular resolution,

\begin{equation}
{\rm DP}_{\lambda_2 /\lambda_1} = {{\rm P}_{\rm nth}^{\lambda_2} \over {\rm P}_{\rm nth}^{\lambda_1}},
\label{eq:DPwav}
\end{equation}
where, $\lambda_2 > \lambda_1$. The observed depolarization ${\rm DP}_{\lambda_2 /\lambda_1}$, that is only wavelength \emph{dependent}, is called the Faraday depolarization.   

We derived the depolarization DP$_{20/3.6}$ using the maps of nonthermal and polarized intensity at 20 and 3.6\,cm at the same angular resolution of 180$\arcsec$ (Fig.~\ref{fig:depol}, top panel).
The southern half of the galaxy is highly depolarized compared to the northern half. While DP$_{20/3.6}$ changes between 0.0 and 0.5 in the south, it varies between 0.3 and 1.0 in the north.  Considerable depolarization is found at the positions of the prominent HII regions NGC604 (RA\,=\,1$^h$ 34$^m$ 32.9$^s$, DEC\,=\,30$^{\circ}$ 47$\arcmin$ 19.6$\arcsec$), NGC595 (RA\,=\,1$^h$ 33$^m$ 32.4$^s$, DEC\,=\,30$^{\circ}$ 41$\arcmin$ 50$\arcsec$) and IC133 (RA\,=\,1$^h$ 33$^m$ 15.3$^s$, DEC\,=\,30$^{\circ}$ 53$\arcmin$ 19.7$\arcsec$) as can be expected due to their high densities of thermal electrons. The strongest depolarization in the inner galaxy occurs in the main southern arm I\,S. No depolarization (DP\,$\simeq$\,1) is seen on the eastern end of the minor axis and some northern regions.

There are several mechanisms that can lead to wavelength-dependent Faraday depolarization \citep{Burn, Sokoloff_98}. \emph{Differential Faraday rotation} occurs when synchrotron emission originates in a magneto-ionic medium containing a regular magnetic field. The polarization plane of the radiation produced at different depths within the source is rotated over different angles by the Faraday effect and this results in a decrease in the measured degree of polarization. \emph{Faraday dispersion} is depolarization due to fluctuations in the rotation measure within a beam, caused by the turbulent magnetic field and distribution of thermal electrons along the line of sight. When this dispersion is intrinsic to the source, it is called internal Faraday dispersion. In case of a dispersion in an external screen it is called external Faraday dispersion. This depolarization effect may be responsible for the north-south asymmetry in the polarized emission from M33, if an asymmetry in distribution of the foreground magneto-ionic medium exists.  However, as M33 cannot be resolved in the available foreground surveys, like RM  \citep{Johnston} and H$\alpha$ (Wisconsin H$\alpha$ mapper) surveys, we do not discuss this depolarization further. Finally \emph{rotation measure gradients} on the scale of the beam or larger due to systematic variation in the regular magnetic field can  also lead to depolarization. The regular field in M33 is not strong enough nor is the inclination of the galaxy high enough for this effect to be significant \citep[in contrast the highly regular field of the strongly inclined galaxy M31 does produce strong RM gradients, ][]{Fletcher_04}. 

\begin{figure}
\begin{center} 
\resizebox{7cm}{!}{\includegraphics*{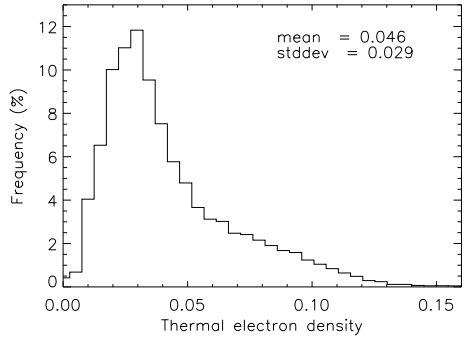}}
\caption{Histogram of the thermal electron density $\langle n_e \rangle$ (cm$^{-3}$) distribution across M33, derived from an extinction corrected H$\alpha$ \citep{Tabatabaei_3_07}. The mean and standard deviation (stddev) of the distribution are given also.}
\label{fig:ne}
\end{center}
\end{figure}
\begin{figure*}
\begin{center}
\resizebox{8cm}{!}{\includegraphics*{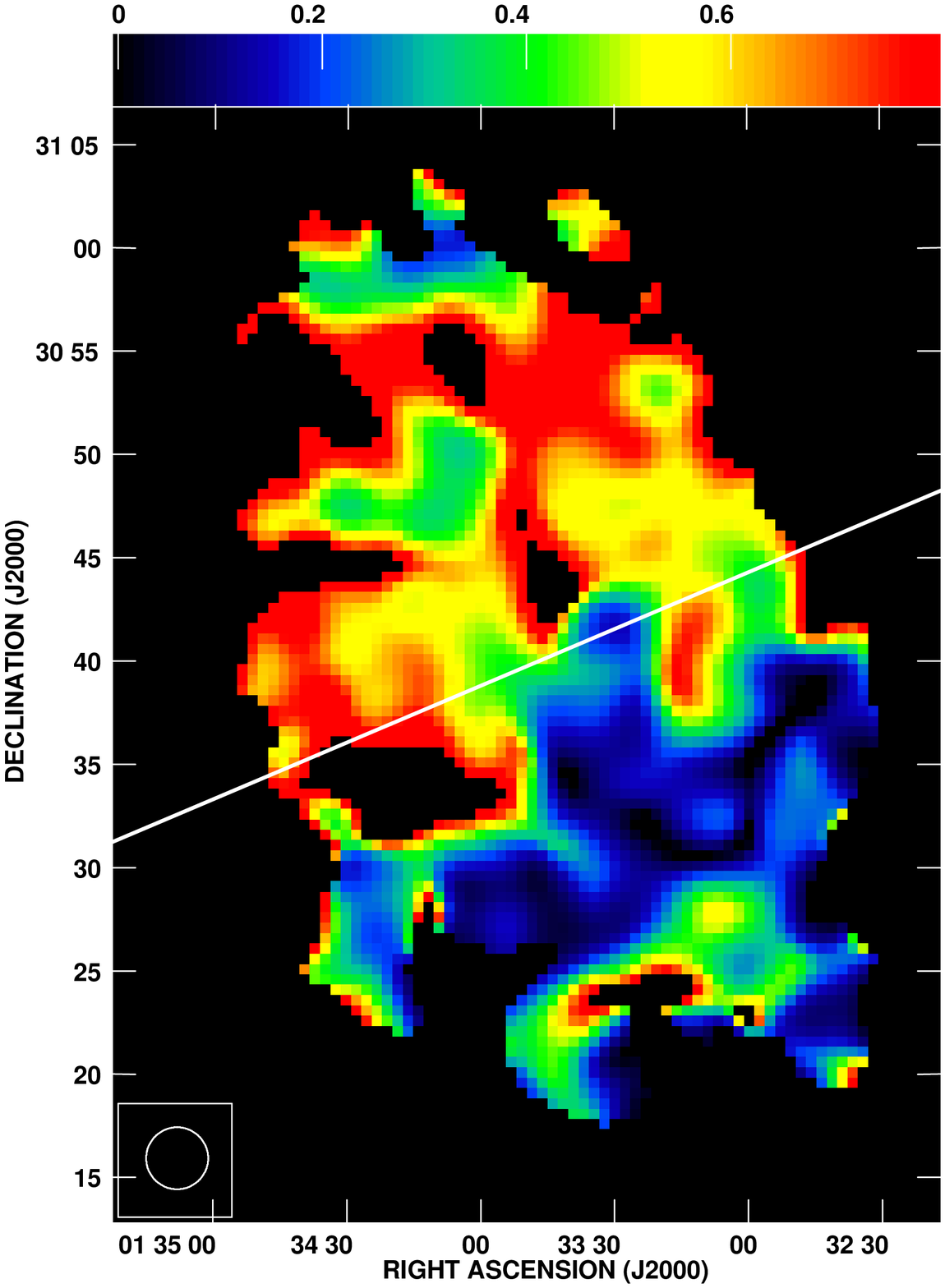}}
\resizebox{\hsize}{!}{\includegraphics*{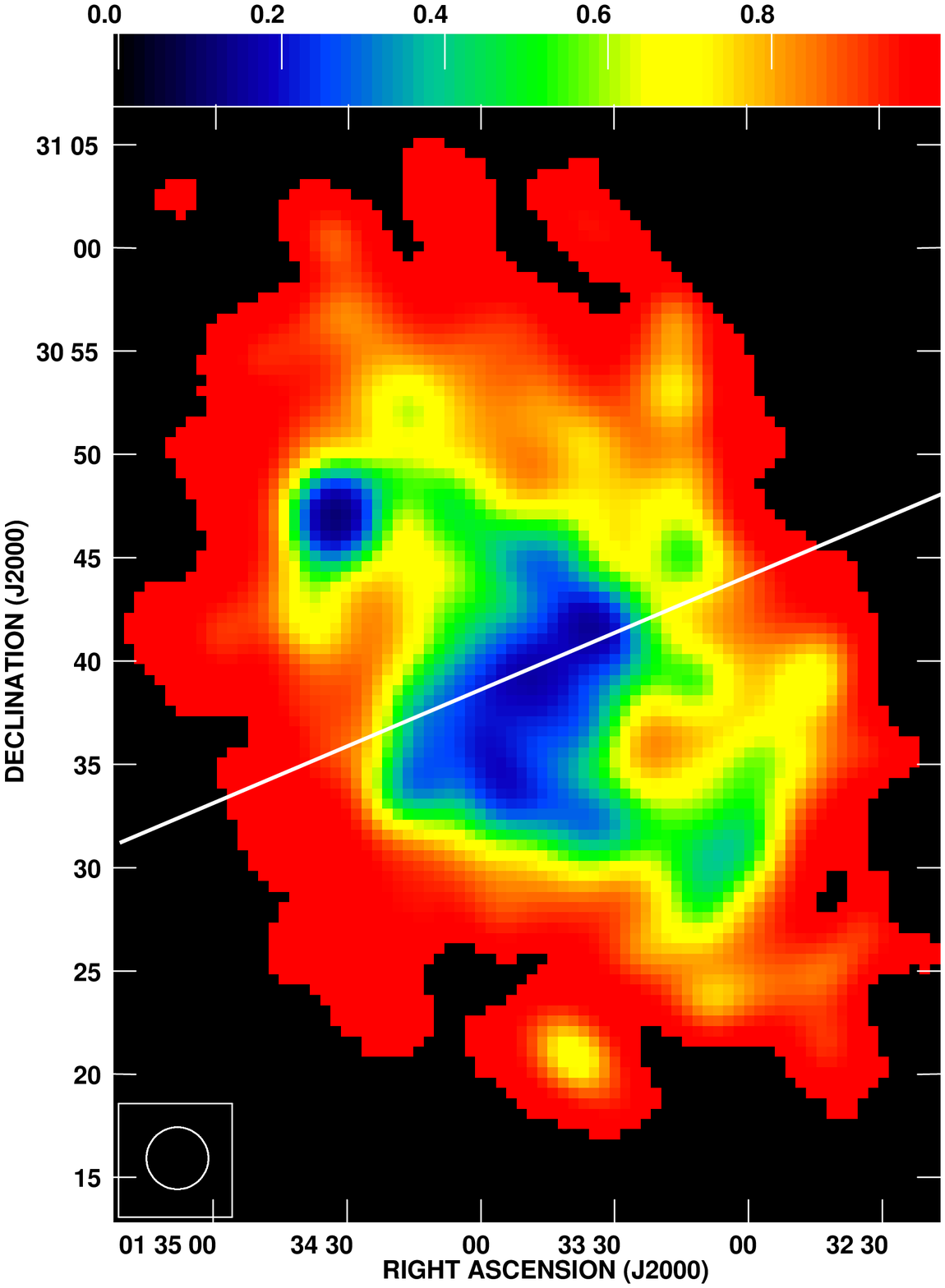}
\includegraphics*{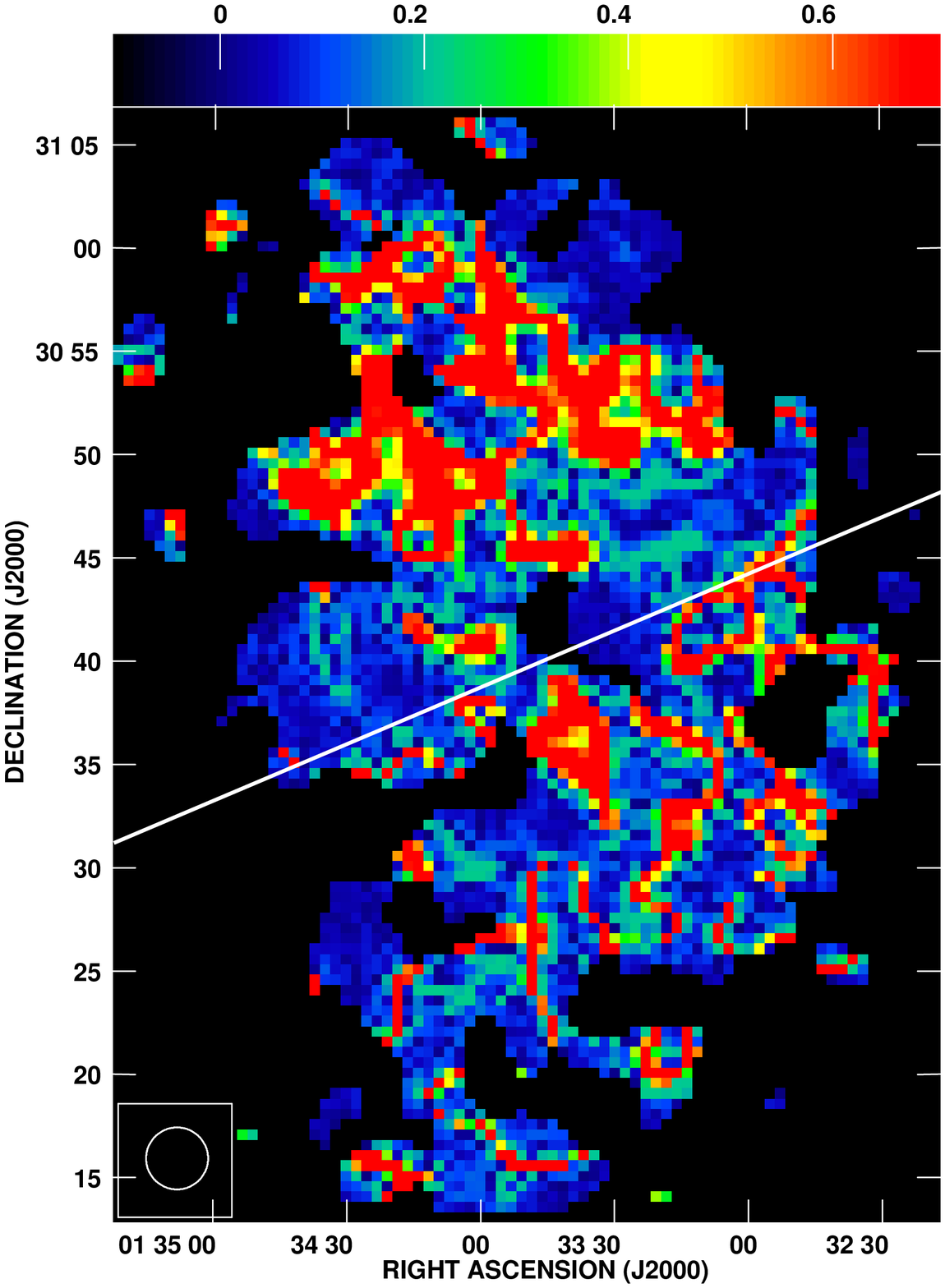}}
\caption{ {\it Top:} observed depolarization ${\rm DP}_{20/3}$ between 3.6 and 20\,cm (Eq.~\ref{eq:DPwav}).  {\it Bottom-left:} estimated depolarization ${\rm DP}_{20/3}$ due to dispersion in Faraday rotation (Eq.~\ref{eq:DPdisp}), and {\it bottom-right:} estimated depolarization ${\rm DP}_{20/3}$ due to  differential Faraday rotation (Eq.~\ref{eq:DPdiff}). The straight line shows the minor axis of M33. }
\label{fig:depol}
\end{center}
\end{figure*}

The depolarization due to internal Faraday dispersion \citep[given by ][]{Sokoloff_98} is 
\begin{eqnarray}
{\rm DP}_{r} & = & \left. {1-e^{-2\sigma_{\rm RM}^2\,\lambda^4} \over 2\sigma_{\rm RM}^2\,\lambda^4}, \right. \nonumber \\
\sigma_{\rm RM} & = & 0.81 \langle n_e \rangle\, \bt\,\sqrt{L\,d\,/\,f}, 
\label{eq:DPdisp}
\end{eqnarray}
where the dispersion in rotation measure is $\sigma_\mathrm{RM}$, with $L$ the pathlength through the ionized medium, $f$ the filling factor of the Faraday-rotating gas along the line of sight \citep[$\simeq$\,0.5, ][]{Beck_07},  and $d$ the turbulent scale \citep[$\simeq$ 50\,pc, ][]{Ohno}.
Using the H$\alpha$ emission measure ($EM = \int{ n_e^2.\,dl}$) and a clumping factor $f_{\rm c} = \,\langle n_e \rangle^2/\langle n_e^2 \rangle$ describing the variations of the electron density, $\langle n_e \rangle$ can be determined by 
$\langle n_e \rangle = \sqrt{{f_{\rm c}\,EM / L}}$.
For the local interstellar medium, \cite{Manchester} found $f_{\rm c}\simeq\,0.05$. Assuming a thickness of $\simeq 1$\,kpc for the thermal electrons in the disk of the galaxy \citep[the  Galactic value, ][]{Cordes} and correcting for the inclination of M33, $L \simeq 1800$\,pc. Then the extinction corrected H$\alpha$ ($EM$) map of M33 \citep{Tabatabaei_3_07} generates a distribution of $\langle n_e\rangle$ across the galaxy with a mean value of $\simeq$ 0.05\,cm$^{-3}$ and a most probable value of $\simeq$ 0.03\,cm$^{-3}$ (Fig.~\ref{fig:ne}), that is in agreement with the estimated values in our galaxy \citep{Cordes} and other nearby galaxies \citep{Krause_1_89,Dumke_00}. Note that a more realistic approach would consider different filling factors and electron densities for the thin  and thick disk of the galaxy. However, because the only information we have is a superposition of these components along the line of sight, we are not able to distinguish the role of each component. 
The resulting $\langle n_e\rangle$ and $\bt$ obtained in Sect.\ref{subsec:equipart} (Fig.~\ref{fig:Br}) enable us to estimate DP$_{r}$ at 3.6 and 20\,cm. 
The left-bottom panel in Fig.~\ref{fig:depol} shows the ratio of DP$_{r}$ at 20 and 3.6\,cm. 

The other Faraday depolarization effect that is strong in M33, differential Faraday rotation, is given by \cite{Burn} and \cite{Sokoloff_98} as,
\begin{equation}
{\rm DP}_u = {\rm sinc}\,\,(2\,{\rm RM}_{i}\, \lambda^2),
\label{eq:DPdiff}
\end{equation} 
where for simplicity we assume that the disk of M33 can be represented as a uniform slab.
Using the RM$_i$ map in Fig.~\ref{fig:rmint}, we estimated DP$_u$ between 20 and 3.6\,cm across the galaxy (Fig.~\ref{fig:depol},  bottom-right). As small variations in RM$_i$ produce large changes in the sinc function in Eq.~(\ref{eq:DPdiff}), the resulting ${\rm DP}_u$ is not smoothly distributed among neighboring pixels.

Qualitatively, both kinds of Faraday depolarization contribute to the observed depolarization in M33. The global phenomenon, the north-south asymmetry, is visible in both DP$_u$ as weaker depolarization (DP$_{20/3}\sim 1$) in the north and also in DP$_r$ as stronger depolarization (DP$_{20/3}\sim 0$) in the south part of the central region.  However, locally e.g. at the positions of HII complexes and the southern spiral arms, DP$_{r}$ could explain the observed depolarization. The contributions of DP$_u$ and DP$_{r}$ vary region by region. 
A more quantitative comparison requires a combination of DP$_u$ and DP$_{r}$ across the galaxy, but this needs a detailed modelling of depolarization along with distribution of the filling factors $f$ and $f_c$, the pathlength $L$, and the turbulent scale $d$ across the galaxy.

In the south of M33, a strong turbulent condition was already indicated from the HI line-widths being larger than in the north \citep{Deul}, which could be connected to the high starformation activities in the southern arms particularly in the main arm I\,S \citep{Tabatabaei_2_07}.
Hence, we conclude that the highly turbulent southern M33 along with a  magneto-ionic medium containing vertical regular magnetic fields, reduce the degree of polarization of the integral emission from the southern half and cause the wavelength-dependent north-south asymmetry in polarization (or depolarization).

\section{Discussion}
\subsection{Vertical magnetic fields}
\label{subsec:Bz}

The model regular magnetic field described in Table~\ref{tab:fit} and shown in Fig.~\ref{fig:polar} has a vertical component in each ring. In the inner ring, the combination of the modes $m=z0$ and stronger $m=z1$ produces a sinusoidal vertical field that is strongest near the major axis: pointing away from us at $\theta\simeq 30\degr$ and towards us at $\theta \simeq 210\degr$. In the outer ring the vertical field is also strongest near the major axis, at $\theta\simeq 10\degr$ (directed away) and $\theta\simeq 190\degr$ (directed towards).

The presence of a vertical component to the regular magnetic field was already indicated locally by our rotation measure maps (Sect.~\ref{sec:RM}). However, the large scale vertical field required by our fits requires a global origin.  The strong $\Buz$ along the major axis together with the large line of sights through the magneto-ionic medium on the eastern and western minor axis (Sect.~\ref{sec:RM}) suggest that warp may play a role.
In other words, the `vertical' field that we identify may be due to the severe warp in M33 \citep{Rogstad,Reakes,Sandage,Corbelli_97}. The inner HI disk investigated by \cite{Rogstad}, shows a warp beginning at a radius of $\simeq$ 5\,kpc with a change in the inclination angle of $40\degr$ at 8\,kpc. The warp of the optical plane begins as close to the center as the first arm system at 2\,kpc (the center of our inner ring), with a change in the arm inclination of $>15\degr$ at 3\,kpc and $25\degr$ at 5\,kpc \citep{Sandage}. The model in Sect.~\ref{subsec:fit} assumes a constant inclination of $i=56\degr$ but in a strongly warped disc $i$ varies with radius and azimuth. In this case, even if $\Br$ only has components in the \emph{warped} disk plane $B\disk$, as for e.g. M51 \citep{Berkhuijsen_97}, M31 \citep{Fletcher_04}, and NGC6946 \citep{Beck_07},  there will be an \emph{apparent} vertical component $\hat{\Buz}$ (as well as an apparent disk parallel component $\hat{B\disk}$) with respect to the \emph{average} disk-plane. 

The ratio $\hat{\Buz}/\hat{B}\disk=\tan i_w$ where $i_w$ is the warp inclination. So for $w_i\simeq 15\degr$ in the inner ring $\hat{\Buz}/\hat{B}\disk\simeq 0.3$ and in the outer ring $w_i\simeq 25\degr$ gives $\hat{\Buz}/\hat{B}\disk\simeq 0.5$. Our model field in Table~\ref{tab:fit} has $\Buz/B\disk\le 2\pm 1$ in the ring $1$--$3$kpc and $\Buz/B\disk\le 1.0 \pm 0.4$ in the ring $3$--$5$kpc. This indicates that,  in the outer ring, the vertical field could be mainly due to the warp. However, a real vertical field of a broadly comparable strength to the disk field can exist in the inner ring.


\subsection{Magnetic and spiral-arm pitch angles}

The pitch angles of the horizontal component of the regular magnetic field are high: $\pb=48\degr$ in the ring 1--3 kpc and $\pb=42\degr$ in the ring 3--5 kpc. These magnetic field pitch angles are however lower than the pitch angles of the optical arm segments identified by \citet{Sandage}, which are typically $\pa=60\degr$--$70\degr$. A combination of shear from the differential rotation, producing an azimuthal magnetic field with $\pb=0\degr$, and compression in spiral arm segments, amplifying the component of the field parallel to the arms $\pb=65\degr$, may be responsible for the observed $\pb\simeq 40\degr$. However, this type of alternate stretching and squeezing of the field could not produce the $m=0$ azimuthal mode that is found in both rings, unless the pre-galactic field was of this configuration. The presence of a significant $m=0$ azimuthal mode of $\Br$ can be explained if a large-scale galactic dynamo  is operating in M33: the axisymmetric mode has the fastest growth rate in disk dynamo models \citep[e.g.][]{Beck_96}. This does not mean that a dynamo is the origin of \emph{all} of the regular magnetic field structure in M33. In particular it would be a strange coincidence if the large $\pb$, higher than the typical $\pb$ in other disc galaxies by a factor of $\sim 2$, is not connected to the open spiral arms with $\pa\simeq 65\degr$.

A rough estimate of the magnetic field pitch angles expected due to a simple mean-field dynamo can be obtained by considering the ratio of the alpha-effect --- parameterizing cyclonic turbulence generating radial field $\Bur$ from azimuthal $\But$ --- to the omega-effect --- describing differential rotation shearing radial field into azimuthal. This can be written as \citep{Shukurov_04}
\begin{equation}
\tan{\pb}=\frac{\Bur}{\But}\simeq\frac{1}{2}\sqrt{\frac{\pi\alpha}{hG}},
\end{equation}
where $\alpha$ is a typical velocity of the helical turbulence, $h$ is the scale height of the dynamo active layer and $G=R\,\mathrm{d}\Omega/\mathrm{d}R$ gives the shear rate due to the angular velocity $\Omega$. Using the HI rotation curve derived by \cite{Corbelli_07},  $\alpha\sim 1\kms$ as a typical value \citep[][]{Ruzmaikin_88}, and an HI scale height of 250pc at $R$=2kpc increasing steadily to 650pc at $R$=5kpc \citep{Baldwin}  we obtain approximate pitch angles of $\pb\simeq 20\degr$ and $\pb\simeq 15\degr$ for the rings 1-3 and 3-5 kpc,  
respectively. 
These are only about 1/2 to 1/3 of the fitted pitch angles  
of the m=0 modes. Models specific to M33, which allow for dynamo action as well as the large scale gas-dynamics of the galaxy, are required to understand the origin of the large $\pb$ as well as the vertical component of the regular magnetic field.

\subsection{Energy densities in the ISM}

\begin{figure}
\begin{center}
\resizebox{8cm}{!}{\includegraphics*{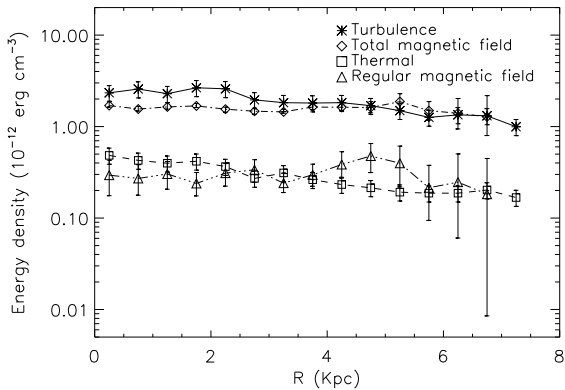}}
\caption{Energy densities and their variations with galactocentric radius in M33.}
\label{fig:energy}
\end{center}
\end{figure}

The energy densities of the equipartition magnetic fields in the disk ($B_{\rm t}^2 /8 \pi$ and $B_{u}^2/8 \pi$ for the total and regular magnetic fields, respectively) are shown in Fig.~\ref{fig:energy}. The thermal energy density of the warm ionized gas, $\frac{3}{2} \langle n_e \rangle kT_e$, is estimated from the H$\alpha$ map assuming $T_e \simeq 10^4$\,K (see also Sect.~\ref{sec:depol}). Assuming the pressure equilibrium between the warm and hot ionized gas with  $T_e \simeq 10^6$\,K and an electron density of $\simeq 0.01 \langle n_e \rangle$ \citep[e.g. ][]{Ferrire}, the energy density of the hot ionized gas is about the same order of magnitude as the warm ionized gas energy density. For the neutral gas, we derive the energy density of $\frac{3}{2} \langle n \rangle kT$ using the average surface density of total (molecular + atomic) gas given by \cite{Corbelli_03} and an average temperature of $T\simeq 50$\,K \citep{Wilson_97}. The warm neutral gas with a typical temperature of $\simeq 6000$\,K has roughly the same thermal energy density as the cold neutral gas, due to a $\simeq100$ times smaller density \citep[e.g. ][]{Ferrire}. Assuming a constant scale height of the disk of 100pc  \citep[as used for NGC6946 ][]{Beck_07}, we obtain a gas density of $\langle n \rangle \simeq$\,6\,cm$^{-3}$ at R=1\,kpc (that is about 8 times smaller than the corresponding value in NGC6946) to $\simeq$\,2\,cm$^{-3}$ at R=5\,kpc (about 3.5 times smaller than that in NGC6946)\footnote{Hence, the radial profile of the energy density of the neutral gas is much flatter in M33 than in NGC6946 \citep[see Fig.~5 in ][]{Beck_07}}.  The total thermal energy density shown in Fig.~\ref{fig:energy} includes the contribution of the warm, hot ionized and the cold, warm neutral gas.  
Figure.~\ref{fig:energy} also shows the kinetic energy density of the turbulent motion of the neutral gas estimated using a turbulent velocity of 10\,km\,s$^{-1}$ \citep{Corbelli_97,Milosavljevi}. 

Generally, the energy densities of all components are about the same order of magnitude as in the Milky Way \citep{Cox_05}, but one order of magnitude smaller than in NGC6946 \citep{Beck_07}. 
The small thermal energy density compared to the total magnetic field energy density shows that the M33's ISM is a low $\beta$ plasma \citep[$\beta$ is defined as the thermal to magnetic energy density ratio, similar results were found in the Milky Way and NGC6946][]{Cox_05,Beck_07}. On the other hand,  the small thermal energy density compared to that of the turbulent motions indicates that turbulence in the diffuse ISM is supersonic \citep[as in NGC6946][]{Beck_07}. This is in agreement with 3-D MHD models for the ISM \citep{deAvillez}. 

The energy densities of the total magnetic field and turbulent gas motions are about the same. This hints to the generation of interstellar magnetic fields from turbulent gas motions for R$<$8\,kpc in M33. For a comparison at larger radii, detection of the magnetic fields is required using deep surveys of Faraday rotation of polarized background sources. This seems to be promising as the scale length of the total magnetic field \citep[$\simeq$24\,kpc, ][]{Tabatabaei_3_07} is much larger than the present detection limit of the radio emission.

The energy density of the regular field is about 3--7 times smaller than that of the total field. Furtheremore, it shows more variations with radius than the total field strength, with a maximum increase at $4.5<R<5.5$\,kpc.

\subsection{Comparison with M31}

Comparing magnetic fields in our neighbors, M33 and M31, is instructive specially because a similar method was used to find out the 3-D magnetic field structure in both galaxies. It is interesting to see that the total magnetic field strength is about the same in M33 and M31 ($\simeq 7\,\mu$G), but the regular field strength in M33 is about half of that in M31 ($\simeq 4\,\mu$G). Furthermore, in contrast to M33, M31 has a disk plane parallel regular field without a vertical component. In other words, the large-scale magnetic field is well-ordered and strong relative to the small-scale field in M31. The regular magnetic field is fitted by a dominant m=0 mode in M31 \citep{Fletcher_04} that is much stronger than that in M33. 

The larger pitch angles of the horizontal magnetic field in M33 than in M31 is not due to a smaller shear in M33. From their rotation curves \citep{Carignan,Corbelli_07}, the shear rate at the relevant radii is larger in M33 ($>10$\,km\,s$^{-1}$\,kpc$^{-1}$ at $1<R<5$\,kpc) than in M31 ($\simeq$\,6\,km\,s$^{-1}$\,kpc$^{-1}$ at $8<R<12$\,kpc). 

The fact that M33 has a higher star formation efficiency than its 10-times more massive neighbor, M31, may be a clue to the origin of their differences.  Strong starformation activities in the inner part of M33 could cause vertical distribution of magneto-ionic matter and hence the vertical magnetic field. Furthermore, stronger turbulence in the interstellar medium can be generally caused by high starformation rate increasing the dynamo alpha-effect and hence providing large pitch angles of the horizontal magnetic field.




\section{Summary}
\label{sec:conc}

The distributions of linearly polarized intensity and polarization angle at 3.6, 6.2, and 20\,cm along with the maps of nonthermal intensity and nonthermal spectral index \citep[obtained from the new separation method, ][]{Tabatabaei_3_07} yielded high-resolution distributions of RM, nonthermal degree of polarization, and Faraday depolarization in M33. Furthermore, we derived the 3-D structure of the regular magnetic field by fitting the observed azimuthal distributions of the polarization angle within two rings of 2\,kpc width in the radial range 1 to 5\,kpc.  The main results and conclusions are as follows:

\begin{itemize}

\item[1. ] The average nonthermal degree of polarization is P$_{\rm nth} \simeq$\,10\% (at 3.6\,cm) for $R<7.5$\,kpc and $>\,20\%$ in parts of the magnetic filament. Due to Faraday depolarization P$_{\rm nth}$ decreases to $\simeq$\,6\% at 20\,cm. \\

\item[2. ] The intrinsic Faraday rotation shows larger small-scale variations and weaker correlation with PI in the south than in the north of M33. The higher starformation activity in the southern arms could increase the turbulent velocities of interstellar clouds and disturb the regular field configuration.  
On the other hand, a good correlation between RM$_i$ and PI in the magnetic filament in the north-west of M33 shows that here the magnetic field is mainly regular.\\

\item[3. ] The average equipartition strengths of the total and regular magnetic fields are  $\Bt\simeq 6.4\,\mu$G  and $\Br\simeq 2.5\,\mu$G for $R<7.5$\,kpc. The regular magnetic field strength is higher within the ring at $4.5<R<5.5$\,kpc, which contains the magnetic filament that has a maximum regular field of $\Br\simeq 6.6\,\mu$G. Strong turbulent magnetic fields  ($\bt>7\,\mu$G) occur in the extended central region and the arms I\,S and part of II\,S.  \\ 

\item[4. ] The 3-D structure of the regular magnetic field can be explained by a combination of azimuthal modes of $m=0+z0+z1$ in the ring 1--3 kpc and $m=0+1+z1$ in the ring 3--5 kpc. The horizontal magnetic field component follows an arm-like pattern with pitch angles smaller than those of the optical arm segments, indicating that large-scale gas-dynamical effects such as compression and shear are not solely responsible for the spiral magnetic lines.
The significant axisymmetric mode (m=0) in both rings indicates that galactic dynamo action is present in M33.\\

\item[5. ] The presence of vertical magnetic fields, shown by the best-fit model of the 3-D field structure ($z1$) and indicated by the Faraday rotation distribution across the galaxy,  is possibly due to both  global (e.g. M33's warp or interaction with M31) and local (e.g.~starformation activities, Parker loops) phenomena. The warp can better explain the origin of the vertical field in the outer ring (3--5\,kpc). \\

\item[6. ] In the southern half of M33, an excess of differential Faraday rotation together with strong Faraday dispersion seem to be responsible for the north-south asymmetry in the observed depolarization (which is wavelength dependent). \\

\item[7. ] The energy densities of the magnetic field and turbulence are about the same, confirming the theory of generation of interstellar magnetic fields from turbulent gas motions. Furthermore, it seems that the ISM in M33 can be characterized by a low $\beta$ plasma and dominated by a supersonic turbulence, as the energy densities of the magnetic field and turbulence are both higher than the thermal energy density.

\end{itemize} 


\begin{acknowledgements}
We are grateful to E. M. Berkhuijsen, U. Klein and E. Kr\"ugel for valuable and stimulating comments. 
FT was supported through a stipend from the Max Planck Institute for Radio Astronomy (MPIfR).  AF thanks the Leverhulme Trust for financial support under research grant F/00 125/N.

\end{acknowledgements}

\bibliography{s.bib}

\begin{thebibliography}{40}
\expandafter\ifx\csname natexlab\endcsname\relax\def\natexlab#1{#1}\fi

\bibitem[{{Baldwin}(1981)}]{Baldwin}
{Baldwin}, J.~E. 1981, in Structure and Evolution of Normal Galaxies, ed. S.~M.
  {Fall} \& D.~{Lynden-Bell}, 137--147

\bibitem[{{Beck}(1979)}]{Beck_79}
{Beck}, R. 1979, Ph.D.~Thesis, Rheinische Friedrich-Wilhelms-Universitaet,
  Bonn.

\bibitem[{{Beck}(2007)}]{Beck_07}
{Beck}, R. 2007, \aap, 470, 539

\bibitem[{{Beck} {et~al.}(1996){Beck}, {Brandenburg}, {Moss}, {Shukurov}, \&
  {Sokoloff}}]{Beck_96}
{Beck}, R., {Brandenburg}, A., {Moss}, D., {Shukurov}, A., \& {Sokoloff}, D.
  1996, \araa, 34, 155

\bibitem[{{Beck} \& {Krause}(2005)}]{Beck_06}
{Beck}, R. \& {Krause}, M. 2005, Astronomische Nachrichten, 326, 414

\bibitem[{{Berkhuijsen} {et~al.}(1997){Berkhuijsen}, {Horellou}, {Krause},
  {Neininger}, {Poezd}, {Shukurov}, \& {Sokoloff}}]{Berkhuijsen_97}
{Berkhuijsen}, E.~M., {Horellou}, C., {Krause}, M., {et~al.} 1997, \aap, 318,
  700

\bibitem[{{Broten} {et~al.}(1988){Broten}, {MacLeod}, \& {Vallee}}]{Broten}
{Broten}, N.~W., {MacLeod}, J.~M., \& {Vallee}, J.~P. 1988, \apss, 141, 303

\bibitem[{{Buczilowski } \& {Beck }(1991)}]{Buczilowski_etal_91}
{Buczilowski }, U.~R. \& {Beck }, R. 1991, \aap, 241, 47

\bibitem[{{Burn}(1966)}]{Burn}
{Burn}, B.~J. 1966, \mnras, 133, 67

\bibitem[{{Carignan} {et~al.}(2006){Carignan}, {Chemin}, {Huchtmeier}, \&
  {Lockman}}]{Carignan}
{Carignan}, C., {Chemin}, L., {Huchtmeier}, W.~K., \& {Lockman}, F.~J. 2006,
  \apjl, 641, L109

\bibitem[{{Corbelli}(2003)}]{Corbelli_03}
{Corbelli}, E. 2003, \mnras, 342, 199

\bibitem[{{Corbelli} \& {Salucci}(2007)}]{Corbelli_07}
{Corbelli}, E. \& {Salucci}, P. 2007, \mnras, 374, 1051

\bibitem[{{Corbelli} \& {Schneider}(1997)}]{Corbelli_97}
{Corbelli}, E. \& {Schneider}, S.~E. 1997, \apj, 479, 244

\bibitem[{{Cordes} \& {Lazio}(2002)}]{Cordes}
{Cordes}, J.~M. \& {Lazio}, T.~J.~W. 2002, ArXiv Astrophysics e-prints

\bibitem[{{Cox}(2005)}]{Cox_05}
{Cox}, D.~P. 2005, \araa, 43, 337

\bibitem[{{de Avillez} \& {Breitschwerdt}(2007)}]{deAvillez}
{de Avillez}, M.~A. \& {Breitschwerdt}, D. 2007, \apjl, 665, L35

\bibitem[{{Deul} \& {van der Hulst}(1987)}]{Deul}
{Deul}, E.~R. \& {van der Hulst}, J.~M. 1987, \aaps, 67, 509

\bibitem[{{Dumke} {et~al.}(2000){Dumke}, {Krause}, \& {Wielebinski}}]{Dumke_00}
{Dumke}, M., {Krause}, M., \& {Wielebinski}, R. 2000, \aap, 355, 512

\bibitem[{{Ferri{\`e}re}(2001)}]{Ferrire}
{Ferri{\`e}re}, K.~M. 2001, Reviews of Modern Physics, 73, 1031

\bibitem[{{Fletcher} {et~al.}(2004){Fletcher}, {Berkhuijsen}, {Beck}, \&
  {Shukurov}}]{Fletcher_04}
{Fletcher}, A., {Berkhuijsen}, E.~M., {Beck}, R., \& {Shukurov}, A. 2004, \aap,
  414, 53

\bibitem[{{Johnston-Hollitt} {et~al.}(2004){Johnston-Hollitt}, {Hollitt}, \&
  {Ekers}}]{Johnston}
{Johnston-Hollitt}, M., {Hollitt}, C.~P., \& {Ekers}, R.~D. 2004, in The
  Magnetized Interstellar Medium, ed. B.~{Uyaniker}, W.~{Reich}, \&
  R.~{Wielebinski}, 13--18

\bibitem[{{Krause}(1990)}]{Krause_90}
{Krause}, M. 1990, in IAU Symposium, Vol. 140, Galactic and Intergalactic
  Magnetic Fields, ed. R.~{Beck}, R.~{Wielebinski}, \& P.~P. {Kronberg},
  187--196

\bibitem[{{Krause} {et~al.}(1989){Krause}, {Hummel}, \& {Beck}}]{Krause_1_89}
{Krause}, M., {Hummel}, E., \& {Beck}, R. 1989, \aap, 217, 4

\bibitem[{{Manchester} \& {Mebold}(1977)}]{Manchester}
{Manchester}, R.~N. \& {Mebold}, U. 1977, \aap, 59, 401

\bibitem[{{Milosavljevi{\'c}}(2004)}]{Milosavljevi}
{Milosavljevi{\'c}}, M. 2004, \apjl, 605, L13

\bibitem[{{Ohno} \& {Shibata}(1993)}]{Ohno}
{Ohno}, H. \& {Shibata}, S. 1993, \mnras, 262, 953

\bibitem[{{Reakes} \& {Newton}(1978)}]{Reakes}
{Reakes}, M.~L. \& {Newton}, K. 1978, \mnras, 185, 277

\bibitem[{{Reich}(2006)}]{Reich_06}
{Reich}, W. 2006, ArXiv Astrophysics e-prints

\bibitem[{{Rogstad} {et~al.}(1976){Rogstad}, {Wright}, \& {Lockhart}}]{Rogstad}
{Rogstad}, D.~H., {Wright}, M.~C.~H., \& {Lockhart}, I.~A. 1976, \apj, 204, 703

\bibitem[{{Ruzmaikin} {et~al.}(1988){Ruzmaikin}, {Sokolov}, \&
  {Shukurov}}]{Ruzmaikin_88}
{Ruzmaikin}, A.~A., {Sokolov}, D.~D., \& {Shukurov}, A.~M., eds. 1988,
  Astrophysics and Space Science Library, Vol. 133, {Magnetic fields of
  galaxies, Chapter VI.4}

\bibitem[{{Sandage} \& {Humphreys}(1980)}]{Sandage}
{Sandage}, A. \& {Humphreys}, R.~M. 1980, \apjl, 236, L1

\bibitem[{{Segalovitz} {et~al.}(1976){Segalovitz}, {Shane}, \& {de
  Bruyn}}]{Segalovitz}
{Segalovitz}, A., {Shane}, W.~W., \& {de Bruyn}, A.~G. 1976, \nat, 264, 222

\bibitem[{{Shukurov}(2004)}]{Shukurov_04}
{Shukurov}, A. 2004, ArXiv Astrophysics e-prints

\bibitem[{{Simard-Normandin} \& {Kronberg}(1980)}]{Simard}
{Simard-Normandin}, M. \& {Kronberg}, P.~P. 1980, \apj, 242, 74

\bibitem[{{Sokoloff} {et~al.}(1998){Sokoloff}, {Bykov}, {Shukurov},
  {Berkhuijsen}, {Beck}, \& {Poezd}}]{Sokoloff_98}
{Sokoloff}, D.~D., {Bykov}, A.~A., {Shukurov}, A., {et~al.} 1998, \mnras, 299,
  189

\bibitem[{{Tabara} \& {Inoue}(1980)}]{Tabara}
{Tabara}, H. \& {Inoue}, M. 1980, \aaps, 39, 379

\bibitem[{{Tabatabaei} {et~al.}(2007{\natexlab{a}}){Tabatabaei}, {Beck},
  {Krause}, {Kr{\"u}gel}, \& {Berkhuijsen}}]{Tabatabaei_4_07}
{Tabatabaei}, F.~S., {Beck}, R., {Krause}, M., {Kr{\"u}gel}, E., \&
  {Berkhuijsen}, E.~M. 2007{\natexlab{a}}, Astronomische Nachrichten, 328, 636

\bibitem[{{Tabatabaei} {et~al.}(2007{\natexlab{b}}){Tabatabaei}, {Beck},
  {Kr{\"u}gel}, {Krause}, {Berkhuijsen}, {Gordon}, \&
  {Menten}}]{Tabatabaei_3_07}
{Tabatabaei}, F.~S., {Beck}, R., {Kr{\"u}gel}, E., {et~al.} 2007{\natexlab{b}},
  \aap, 475, 133

\bibitem[{{Tabatabaei} {et~al.}(2007{\natexlab{c}}){Tabatabaei}, {Krause}, \&
  {Beck}}]{Tabatabaei_2_07}
{Tabatabaei}, F.~S., {Krause}, M., \& {Beck}, R. 2007{\natexlab{c}}, \aap, 472,
  785

\bibitem[{{Wilson} {et~al.}(1997){Wilson}, {Walker}, \& {Thornley}}]{Wilson_97}
{Wilson}, C.~D., {Walker}, C.~E., \& {Thornley}, M.~D. 1997, \apj, 483, 210

\end{thebibliography}

\end{document}